\documentclass[a4paper,fleqn,usenatbib]{mnras}

\usepackage{newtxtext,newtxmath}

\usepackage[T1]{fontenc}
\usepackage{ae,aecompl}

\usepackage[flushleft]{threeparttable}
\usepackage{pdfpages}

\usepackage{graphicx}	
\usepackage{amsmath}	
\usepackage{amssymb}	

\title[Benchmark ages for the \textit{Gaia} benchmark stars]{Benchmark ages for the \textit{Gaia} benchmark stars}

\author[C. L. Sahlholdt et al.]{
Christian L. Sahlholdt$^{1}$\thanks{E-mail: sahlholdt@astro.lu.se},
Sofia Feltzing$^{1}$,
Lennart Lindegren$^{1}$,
and Ross P. Church$^{1}$
\\
$^{1}$Lund Observatory, Department of Astronomy and Theoretical Physics, Box 43,
      SE-221 00 Lund, Sweden\\
}

\date{Accepted XXX. Received YYY; in original form ZZZ}

\pubyear{2018}

\begin{document}
\label{firstpage}
\pagerange{\pageref{firstpage}--\pageref{lastpage}}
\maketitle

\begin{abstract}
In the era of large-scale surveys of stars in the Milky Way, stellar ages are crucial for studying the evolution of the Galaxy.
But determining ages of field stars is notoriously difficult; therefore, we attempt to determine benchmark ages for the  extensively studied \textit{Gaia} benchmark stars which can be used for validation purposes.
By searching the literature for age estimates from different methods and deriving new ages based on Bayesian isochrone fitting, we are able to put reliable limits on the ages of 16 out of the 33 benchmark stars.
The giants with well-defined ages are all young, and an expansion of the sample to include older giants with asteroseismic ages would be beneficial.
Some of the stars have surface parameters inconsistent with isochrones younger than 16~Gyr.
Including $\alpha$-enhancement in the models when relevant resolves some of these cases, but others clearly highlight discrepancies between the models and observations.
We test the impact of atomic diffusion on the age estimates by fitting to the actual surface metallicity of the models instead of the initial value and find that the effect is negligible except for a single turn-off star.
Finally, we show that our ability to determine isochrone-based ages for large spectroscopic surveys largely mirrors our ability to determine ages for these benchmark stars, except for stars with $\log g \gtrsim 4.4$~dex since their location in the HR~diagram is almost age insensitive.
Hence, isochrone fitting does not constrain their ages given the typical uncertainties of spectroscopic stellar parameters.
\end{abstract}




\section{Introduction}
We have entered a new era of Galactic Archaeology thanks to the wealth of kinematical and chemical information coming from large-scale Galactic surveys.
The \textit{Gaia} spacecraft \citep{2016A&A...595A...1G} measures positions, proper motions, and parallaxes for stars across the Milky Way and the recent second data release \citep{2018A&A...616A...1G} contains astrometric and photometric data for 1.3 billion stars.
At the same time, ground-based spectroscopic surveys are providing stellar parameters like metallicity, effective temperature, surface gravity, and abundances of individual elements.
For example, the GALactic Archaeology with HERMES (GALAH; \citealt{2015MNRAS.449.2604D}) survey recently published spectroscopically derived stellar parameters for about $340\,000$ stars in their second data release \citep{2018MNRAS.tmp.1218B}.
With the combined data from \textit{Gaia} and ground-based spectroscopic surveys, the number of stars in the Milky Way with both kinematical and chemical information is reaching into the millions, allowing for detailed studies of the chemo-kinematical structure of our Galaxy.

In order to put the chemo-kinematical information into an evolutionary context, and thereby learn about the formation history of Galactic structures, we need precise and accurate age estimates for individual field stars.
Unfortunately, age determination of field stars is far from straightforward; our ability to determine a star's age depends on its evolutionary stage, and no single method works well for all stars \citep{2010ARA&A..48..581S}.
Recently, methods have been developed to derive masses, and implied ages, of giants based on their surface carbon and nitrogen abundances \citep{2015MNRAS.453.1855M, 2016MNRAS.456.3655M}.
The relationship between surface abundances and mass can be calibrated using stars with precise mass estimates based on asteroseismology and then applied to large samples of giants.
This method has been applied to estimate ages for more than $200\,000$ giants based on their observed spectra \citep[e.g.][]{2017ApJ...841...40H}.
When analysing such large samples of stars, it is easy to overlook spurious trends and small biases in the derived ages which could be interpreted as real signals.
With this in mind, it would be useful to have a set of calibration/verification stars with well-known ages.

The \textit{Gaia} FGK benchmark stars \citep{2014A&A...564A.133J, 2015A&A...582A..49H} are a sample of 33 nearby and extensively studied stars (34 when including the Sun) which are to be used for calibration of the stellar parameters derived from \textit{Gaia} data (using the \textit{Gaia} astrophysical parameters inference system; \citealt{2013A&A...559A..74B}).
Therefore, they have been chosen to span most of the Hertzsprung-Russell (HR) diagram and sample both solar and sub-solar metallicities.
Their effective temperatures and surface gravities have been determined by \citet{2015A&A...582A..49H} independently of spectroscopy by using angular diameters and bolometric fluxes.
This makes them suitable for validation of stellar parameters from large spectroscopic surveys.
In fact, they have been used for just that purpose in the second GALAH data release to validate the stellar temperatures, metallicities, and surface gravities.

In this paper we investigate the ages of the \textit{Gaia} benchmark stars and attempt to define benchmark ages which can be used to verify age determinations from different methods and pipelines.
By searching the literature for age estimates of these stars, we investigate how well their ages can be constrained using different methods, including those available to large spectroscopic surveys.
At the same time we use a recently developed Bayesian isochrone fitting tool to derive age estimates for the benchmark stars.
This allows us to test what can be achieved using this method with upcoming spectroscopic data, in the best-case scenario, for different stellar types.
We adopt the benchmark stellar parameters for the isochrone-based age determinations which can be considered a snap-shot in time of what is known about the stellar parameters for the \textit{Gaia} Benchmark stars.
We are aware of the on-going work to improve these measurements for many of the stars: for example HD~140283, 122563, and 103095 have had new, improved interferometric measurements leading to updates of the derived effective temperatures \citep{2018MNRAS.475L..81K}.

This paper is organized as follows.
In Section~\ref{sec:lit_ages} the age estimates found in the literature are presented, and the different age-dating methods are discussed.
The data, method, and models used to determine ages from Bayesian isochrone fitting are introduced in Section~\ref{sec:bayes_ages} where we also discuss the typical shapes and information content of the age probability distribution functions for stars of different spectral classes.
Based on the ages found in the literature and determined in this work, we give recommended benchmark ages in Section~\ref{sec:results}, and in Section~\ref{sec:discussion} we discuss their reliability and compare the isochrone-based ages with the benchmark values with an outlook towards age determination of stars in large spectroscopic surveys.
Finally, the conclusions are given in Section~\ref{sec:conclusions}.

\section{Ages in the literature} \label{sec:lit_ages}
\subsection{Search strategy}
In order to find age estimates in the literature, we first searched the SIMBAD database for all publications which make reference to each of the stars and picked out those with the word ``age'' in the abstract.
We chose to limit the scope of the search to going back to the year 1997 since this was when the \textit{Hipparcos} catalogue \citep{1997ESASP1200.....E} was released which improved the stellar luminosity estimates and thus also the age estimates based on isochrone fitting.
We make sure to only take ages from the original source; however, we do include different age estimates which are based on the same stellar data and/or method.
This means that some of the ages we find are correlated, but they help give a better impression of the systematic uncertainties related to the use of different data or models.
We also note the method used to determine each of the ages we find.
If the age is based on model fitting, we note which models were used and the input stellar parameters used to constrain the models.
This makes it easier to understand the differences between the age estimates; an outlier among the ages based on model fitting is usually due to questionable stellar parameters.

By only looking through papers with the word ``age'' in the abstract we are bound to miss some of the age determinations available in the literature.
Along the way we found ages in papers without ``age'' in the abstract simply by following references from other papers.
These ages were added to the final compilation.
In the end, the number of such discoveries was low which leads us to believe that we have found the majority of the age determinations available in the literature for these stars.
We find a varying number of literature ages for each star with 33 values for 18~Sco being the highest.
For every star at least one value was found in the literature, and, in general, many values were found for the dwarfs and subgiants and few for the giants.
The complete collection of literature ages is given in \autoref{tbl:lit_ages} and is, for each star, graphically displayed in Appendix~\ref{sec:appA}.

\begin{table*}
  \centering
  \caption{Ages compiled from the literature.
  Uncertainties in the form of lower and upper confidence limits (as given in the original sources) are given, when available, in the columns ``Min age'' and ``Max age''.
  The effective temperature and/or metallicity used to derive the ages from model fitting (including asteroseismology) are also given here when available.
  For a short discussion of the different methods, see Section~\ref{sec:method_discussion}.
  All of the ages are also graphically displayed in Appendix~\ref{sec:appA}.
  The complete table is available online.}
  \label{tbl:lit_ages}
  \begin{tabular}{lrrrrrrrrll}
    \hline
    Name        & HD  & HIP & Age [Gyr]   & Min age & Max age & $T_{\mathrm{eff}}$ [K] & [Fe/H]   &
    Method   & Source       \\ \hline
    Procyon  & 61421  & 37279 & $1.48$         & $0.73$   & $3.45$    & $6652$               & ---       &
    Model fitting    & {\citet{2015ApJ...804..146D}}   \\
    Procyon  & 61421  & 37279 & $1.85$          & $1.77$    & $1.93$     & ---                & ---        &
    Asteroseismology & {\citet{2014ApJ...787..164G}}       \\
    Procyon  & 61421  & 37279 & $2.47$          & $2.34$    & $2.60$      & ---                & ---        &
    Asteroseismology & {\citet{2014ApJ...787..164G}}        \\
    Procyon  & 61421  & 37279 & $2.80$           & $2.10$     & $3.50$      & $6494$               & $0.02$       &
    Asteroseismology & {\citet{2014A&A...566A..82L}}       \\
    Procyon  & 61421  & 37279 & $2.44$          & $1.91$    & $2.97$     & ---                & ---        &
    Model fitting    & {\citet{2014MNRAS.443..698S}} \\
    \dots \\
    61 Cyg B & 201092 & 104217 & $6.00$             & $5.00$       & $7.00$        & $4040$               & $-0.27$      &
    Model fitting    & {\citet{2008A&A...488..667K}}        \\
    61 Cyg B & 201092 & 104217 & $3.39$          & ---     & ---      & ---                   & ---           &
    Gyrochronology   & {\citet{2008ApJ...687.1264M}} \\
    61 Cyg B & 201092 & 104217 & $3.75$          & ---     & ---      & ---                   & ---           &
    Chromochronology & {\citet{2007ApJ...669.1167B}}                 \\
    61 Cyg B & 201092 & 104217 & $1.87$          & $1.57$    & $2.17$     & ---                   & ---           &
    Gyrochronology   & {\citet{2007ApJ...669.1167B}}                 \\
    61 Cyg B & 201092 & 104217 & ---           & ---     & $0.68$     & ---                   & ---           &
    Model fitting    & {\citet{2007ApJS..168..297T}}         \\ \hline
  \end{tabular}
\end{table*}

\subsection{Discussion of age determination methods} \label{sec:method_discussion}
Many of the stars have had their ages estimated with different methods.
These methods can be divided into two main groups: those based on fitting stellar models to observed stellar parameters and those based on the stellar rotation period or activity level.
A brief discussion of these methods is given here, including their strengths and weaknesses, and some of the key literature references are highlighted for further reading.

\subsubsection{Fitting to stellar models} \label{sec:lit_fitting}
The majority of the compiled literature ages are based on fitting the stellar parameters to a set of stellar models, and at least one of these age estimates has been found for all of the benchmark stars.
In the following, the stellar parameters used as input to the fit are referred to as ``observables'' even when they are derived quantities (based e.g. on the observed spectrum).
This is simply to distinguish them from the quantities derived from the fit to stellar models (e.g. the age).
These ages are based on stellar models computed with a wide variety of stellar evolution codes, a number of different combinations of observables, and many different fitting algorithms.
The most common combination of observables used in the literature is the surface metallicity, effective temperature, and absolute magnitude or luminosity \citep{2005ApJS..159..141V, 2007PASJ...59..335T, 2011A&A...530A.138C}.
The latter of these is based on the observed apparent magnitude and the parallax.
Other applied observables include the surface gravity \citep{2014A&A...562A..71B} or a colour index instead of the temperature \citep{2002A&A...394..927I}.
In some cases the stellar radius derived from interferometric observations has been included in the fit (e.g. \citealt{2008A&A...488..667K}).

Common to all of these estimates is the use of a grid of stellar models with different metallicities, initial masses, and ages, from which the age is estimated by finding the model which best matches the observed stellar parameters.
However, finding this best-fitting model is not straightforward due to the complex shapes of stellar evolutionary tracks/isochrones.
In some regions of the HR~diagram, isochrones of different ages cross over each other meaning that the observed stellar surface parameters may fit equally well to a number of different ages.
Additionally, as pointed out by \citet{2004MNRAS.351..487P}, age estimates based on the best-fitting model do not take into account the fact that some regions of the HR~diagram are inherently more densely populated than others due to, for example,  differences in the evolutionary time scales and the stellar initial mass function.
In order to take these effects into account and avoid a biased age estimate, one can apply a Bayesian fitting algorithm as described by \citet{2004MNRAS.351..487P} and \citet{2005A&A...436..127J}.
The Bayesian method gives the age in terms of a probability density function which gives a more nuanced picture of the stellar age than the most likely value.
Today, Bayesian fitting algorithms are widely used for stellar parameter estimation.
A variant of the Bayesian method, first described in \citet{2006A&A...458..609D}, is used in the PARAM\footnote{Web interface: \url{http://stev.oapd.inaf.it/cgi-bin/param}} code to which we have found many references in our search for ages.
For the subgiant $\eta$~Boo, the transition from traditional to Bayesian model fitting is seen directly in the literature ages.
For this star, the oldest age estimates based on the best-fitting model average close to 3~Gyr, whereas more recent Bayesian estimates, starting with the ages derived for the Geneva-Copenhagen Survey (GCS; \citealt{2004A&A...418..989N}), average close to 2~Gyr.
The ambiguity in the age arises due to $\eta$~Boo's placement in the HR~diagram where both a 2~Gyr main sequence model and a 3~Gyr model at the hook can describe the stellar surface parameters.
The Bayesian algorithms favour the main sequence model since this is the slower phase of evolution.

The single source from which we have obtained the most age estimates is \citet{2014MNRAS.443..698S} who give ages for 30 of the benchmark stars.
They also apply a Bayesian algorithm; however, it stands out among all of the other algorithms in the way they constrain their derived stellar parameters.
Instead of using a set of observed stellar surface parameters derived from spectroscopy, they make use of the stellar spectra directly and fit them to model atmospheres.
The spectra, together with photometric and astrometric information, allow them to constrain their core parameters of effective temperature, metallicity, and surface gravity.
These core parameters then constrain other stellar parameters, including the age, through a set of stellar models.
This method differs from the more commonly applied Bayesian algorithms by allowing a non-Gaussian PDF in the spectroscopic parameters, and \citeauthor{2014MNRAS.443..698S} argue that this is vital for unbiased parameter estimates.

Regardless of the adopted stellar parameters and fitting algorithms, for some stars reliable ages cannot be estimated based on model fitting.
\autoref{fig:iso_plot} shows a number of isochrones which are discussed in greater detail in Section~\ref{sec:isochrones}.
It is clear that isochrones of different ages only separate clearly in the HR~diagram around the turn-off and subgiant branch.
On the lower main sequence the isochrones converge which simply means that the stellar surface parameters carry no useful age information.
The isochrones also converge towards the giant branch, although to a lesser degree than on the main sequence; the youngest isochrones ($\lesssim 2$~Gyr) are separated from the older ones.
What this means is that the estimation of reliable ages of field stars based on model fitting is limited to certain regions in the HR~diagram where the stellar surface parameters are changing with age.

In special cases, namely when a star shows solar-like oscillations, it is possible to constrain its inner structure which does provide age information, even for low-mass main sequence stars and red giants (see \citealt{2013ARA&A..51..353C} for a review of asteroseismology).
We have found age estimates based on model fitting with asteroseismic constraints for nine of the benchmark stars and these have been put in a separate category from the rest of the model fits due to the unique constraints imposed by asteroseismic observables.

Even when an age based on stellar model fitting is well-determined in the sense that its statistical uncertainty is low, one must keep in mind that systematic errors are introduced by uncertainties in the stellar models.
Different evolutionary codes may apply different chemical compositions, fundamental physics (e.g. equation of state, opacity of the stellar matter, and nuclear reaction rates), and physical processes (e.g convective overshooting, microscopic diffusion, and mass loss).
In current one-dimensional stellar models, convective energy transport is typically parametrised in terms of an effective mixing length \citep{1958ZA.....46..108B} which is calibrated such that a model of solar mass and metallicity matches the solar parameters at its current age.
It has been shown, based on 3D convection simulations, that the mixing length varies with stellar parameters \citep{2014MNRAS.445.4366T}.
Therefore, the adoption of a solar calibrated mixing length may introduce systematic errors in the stellar models of giant stars.
Additionally, different treatments of the surface boundary conditions of the models can introduce temperature shifts of up to 100~K on the giant branch \citep{2018A&A...612A..68S, 2018ApJ...860..131C}.
These additional caveats are worth keeping in mind when considering the model fitting ages of giants.

\subsubsection{Rotation/activity-age relations}
Low-mass main-sequence stars have convective surface layers, and the combination of rotation and convective motion can sustain a magnetic field through the dynamo effect.
The magnetic field causes activity in the chromosphere and greatly increases the loss of angular momentum through stellar winds.
Therefore, the star spins down with time making the rotation period a function of age and mass.
The procedure of inverting this relation to determine the age based on the rotation period is known as gyrochronology \citep{2003ApJ...586..464B}.
Additionally, as the rate of rotation decreases, so does the magnetic activity.
Thus, there is also a relationship between the stellar age and activity level.
There are different ways to quantify the stellar activity, and the most common one -- which is also the most frequent in our literature search -- is based on chromospheric emission seen in the cores of the Ca II H and K absorption lines, where the contamination from the photospheric emission is minimised.
This age-dating method is referred to as chromochronology.
Our biggest sources of gyro-/chromochronology ages are \citet{2007ApJ...669.1167B} who also discusses the viability of gyrochronology in comparison with other age determinations, \citet{2004ApJS..152..261W} who present and analyse their own activity measurements, and \citet{2008ApJ...687.1264M} who presented the most extensive calibration of the methods to date.
Most of the recent age estimates we have found based on rotation or activity apply the calibrations by \citet{2008ApJ...687.1264M}.
Ages published earlier (e.g. \citealt{2004ApJS..152..261W, 2007ApJ...669.1167B}) mainly used the calibration by \citet{1993PhDT.........3D, 1998ASPC..154.1235D} for activity-based ages.
Another measure, which is less commonly applied, is the total X-ray luminosity as a fraction of the bolometric luminosity.
Among the few X-ray based ages we have found in the literature, most of them are based on a conversion of the X-ray activity index into the chromoshperic activity index before applying the chromochronology relation (e.g. \citealt{2012AJ....143..135V}).

These rotation/activity-age relations allow for more reliable age estimates for low-mass main-sequence stars than are possible by fitting to stellar evolutionary models.
Whereas we have found at least one age estimate for each star based on model fitting (since it can be applied even when it yields little to no information), we have found rotation/activity-based ages for 15 of the benchmark stars, all of them dwarfs or subgiants.
However, these ages have their own limitations, and they are not necessarily equally reliable for all of
the stars.
One of the limitations is that the relations are mainly calibrated to isochrone ages of nearby clusters, which are young.
This means that the relations are poorly calibrated for ages beyond that of the Sun.
In recent years the samples of cluster stars with measured rotation periods have been expanded to include stars in the 2.5~Gyr old cluster NGC~6819 \citep{2015Natur.517..589M} and in the 4~Gyr old cluster M67 \citep{2016ApJ...823...16B}.
These studies have confirmed that rotation is indeed a good age indicator (with precision of ~17 per cent for stars similar to those in M67) up to at least the solar age, and possibly all the way until the turn-off.
\citet{2016ApJ...823...16B} found that the rotational and chromospheric ages of the individual stars in M67 give the same mean age of the cluster, but the chromospheric ages have a standard deviation of 38 per cent of the mean value compared to 17 per cent for the rotational ones.
This is in line with previous studies that found that the precision of chromochronology for individual stars is
about 40--60 per cent \citep{2008ApJ...687.1264M, 2010ARA&A..48..581S}.
So activity is a less precise age indicator than rotation which is because it is a secondary effect, and stars with the same rotation period will show a range of activity levels due to variability of the activity on multiple time scales.
For example, the age of the Sun as derived from chromochronology changes by about 2~Gyr from minimum to maximum activity during the 11-year solar cycle \citep{2012AJ....143..135V}.
The X-ray luminosity has been shown to have nearly the same age-dating potential as chromospheric activity \citep{2008ApJ...687.1264M}; however, the X-ray luminosity is more variable which increases the uncertainty of single age estimates \citep{2010ARA&A..48..581S}.

In order to calibrate gyrochronology at higher ages, there have been efforts to include field stars with ages based on asteroseismology.
\citet{2015MNRAS.450.1787A} calibrated their gyrochronology relation using, in addition to a couple of clusters, a few hundred stars with ages derived from model fitting including asteroseismic constraints.
They give ages for three of the benchmark stars (18~Sco and $\alpha$~Cen~A \& B) based on their own calibration and the calibrations by \citet{2007ApJ...669.1167B} and \citet{2008ApJ...687.1264M}.
The three ages agree well in each case.
However, they also note that no single model was able to fit all of their cluster and asteroseismic data simultaneously, but it is not clear whether this is a problem with the model or the data.
\citet{2016Natur.529..181V} proposed that older dwarf stars are subject to a weakened magnetic braking which causes the old field stars to be poorly described by the gyrochronology relations calibrated to young clusters.
If this is true, the cluster calibrations of gyrochronology will underestimate the ages of older dwarf stars which have entered the phase of weakened braking.

\begin{figure}
  \center
  \includegraphics[width=\columnwidth]{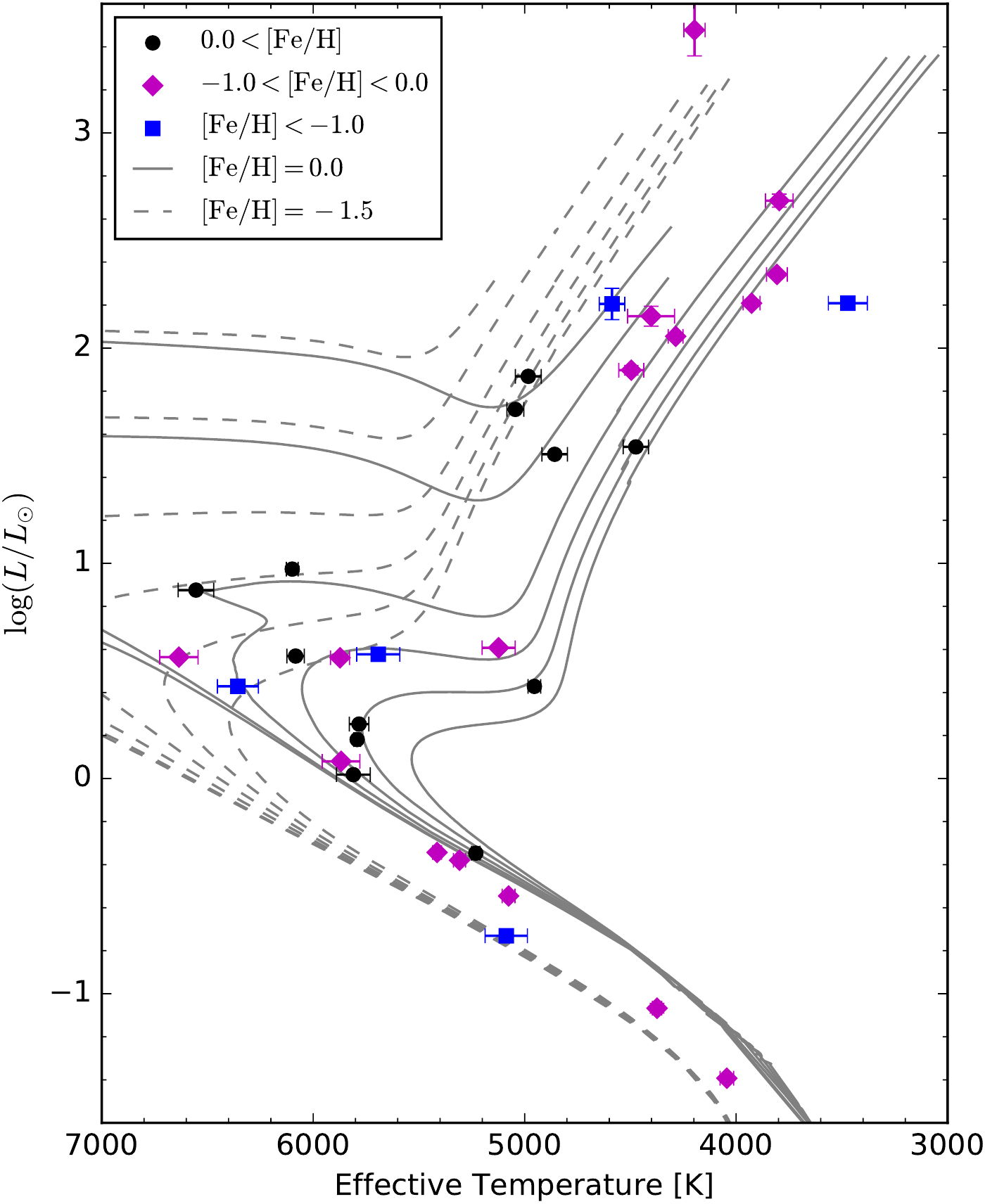}
  \caption{HR~diagram of the \textit{Gaia} benchmark sample.
  The points show the adopted stellar parameters for the stars, and the lines show MIST isochrones with metallicities of ${\mathrm{[Fe/H]} = 0}$ (solid) and ${\mathrm{[Fe/H]} = -1.5}$ (dashed) and for ages of 0.5, 1, 3, 6, 10, and 15~Gyr. The isochrones are only shown for evolutionary stages up to the end of the red giant branch.}
  \label{fig:iso_plot_sample}
\end{figure}

\section{Ages from Bayesian isochrone fitting} \label{sec:bayes_ages}

\subsection{Observational data} \label{sec:obs_data}
In this work we will derive two sets of ages for each star based on model fitting.
The first set is based on fitting to stellar parameters which can all be obtained from spectroscopic data: the effective temperature, $T_{\mathrm{eff}}$, the metallicity, [Fe/H], and the surface gravity, $\log g$.
These will be referred to as $\log g$-based ages.
For the second set $T_{\mathrm{eff}}$ and [Fe/H] is also used, but instead of $\log g$, the parallax, $\varpi$, and the apparent $V$-band magnitude, $V$, are included.
The combination of $\varpi$ and $V$ constrains the absolute magnitude of the star, $M_V$. 
These ages will be referred to as magnitude-based.
We have chosen to test both of these sets of observables to see how well ages can be determined from spectroscopic data alone, compared to what is possible with parallaxes.
We describe the method used to determine ages based on these observables in Section~\ref{sec:fit_summary}.

For all stellar parameters, we adopt the values determined in the original benchmark studies \citep{2015A&A...582A..49H, 2014A&A...564A.133J}.
This means that we mainly take $T_{\mathrm{eff}}$ and $\log g$, and their uncertainties, from Table~10 in \citet{2015A&A...582A..49H}; however, for some stars $T_{\mathrm{eff}}$ and/or $\log g$ are not recommended for use as reference values.
In some of these cases we adopt different values (e.g. spectroscopic values from the literature), and the details of this are given for each star individually in Appendix~\ref{sec:appA}.
We take all values of [Fe/H] from Table~1 in \citet{2015A&A...582A..49H} where the uncertainties are also given based on the combination of all the different uncertainty terms given in Table~3 in \citet{2014A&A...564A.133J}.
We refer to the adopted values of $T_{\mathrm{eff}}$, $\log g$, and [Fe/H] as the benchmark values even though literature values have been adopted in a few cases for $T_{\mathrm{eff}}$ and $\log g$.

For the stellar parallaxes, we adopt the same values as used in \citet{2015A&A...582A..49H}, taken from their Table~7.
These parallaxes come from the revised \textit{Hipparcos} catalogue \citep{2007ASSL..350.....V} for most stars, from \citet{1999A&A...341..121S} for $\alpha$~Cen~A \& B, and from \citet{2014ApJ...792..110V} for the two stars HD~84937 and HD~140283.
Although 22 of the benchmark stars have parallaxes in \textit{Gaia} DR2 \citep{2018A&A...616A...1G}, we have chosen not to use them in this work as they are not necessarily of higher quality than the adopted values in \autoref{tbl:obs_data}, which mainly come from \textit{Hipparcos}.
Only 11 of them have formal parallax uncertainties in DR2 smaller than in \autoref{tbl:obs_data}.
Of these, all but three are brighter than $G\simeq 6$, for which the DR2 parallaxes are known to be unreliable \citep[e.g.,][]{2018ApJ...861..126R}.
The remaining three stars (HD~22879, 103095, and 140283) have DR2 parallaxes that are about 1~mas smaller than in \autoref{tbl:obs_data}, casting doubt on their reliability as well.
Thus, in this work we have deemed it safest not to use the DR2 parallaxes for any of the benchmark stars.

Finally, we have collected $V$-band magnitudes for the stars from the SIMBAD database.
In the cases where no uncertainty is given on the magnitude, we have set it to 0.02~mag.
All of the adopted stellar parameters, as well as their uncertainties, are listed in \autoref{tbl:obs_data} and the sample is visualised in an HR~diagram in \autoref{fig:iso_plot_sample}.

In principle, the $V$~magnitudes should be corrected for interstellar reddening ($E(B-V)$) before they are applied to determine stellar ages.
However, all of the \textit{Gaia} benchmark stars have large parallaxes, i.e. they are nearby.
All but four of the stars are within 100~pc, and it is well known that the Sun is in a Local Bubble where the interstellar reddening is essentially zero \citep[e.g.][]{2003A&A...411..447L}.
Following e.g. \citet{2007AJ....133.2464L}, we find that it is safe to set the reddening for the \textit{Gaia} benchmark stars to zero.
For the four most distant stars (HD~122563, HD~220009, $\beta$~Ara, and $\psi$~Phe), we have performed extra checks and consulted the recent literature.
One of the stars has reddening ($E(BP-RP)$) determined from \textit{Gaia} data (HD~220009, \textit{Gaia} DR2 source 2661005953843811456).
The extinction for this star is ${A_{\mathrm{G}}=0.24}$.
The accuracy of this extinction value is questionable given that it has been derived using the \textit{Gaia} DR2 parallax which may be biased (as discussed above) and which is 2~mas higher than the \textit{Hipparcos} value.
The higher parallax places the star closer to the Sun and introduces the need for extinction to explain the observed $G$~magnitude.
Additionally, even if the \textit{Gaia} parallax is accurate, the \textit{Gaia} stellar parameters suffer from strong systematics in some cases due to the assumptions made in their derivation \citep{2018A&A...616A...8A}.
The three other stars do not have reddening determined in \textit{Gaia} DR2.
Two of them have reddening estimates in the literature: HD~122563 has ${E(B-V) = 0.025}$ \citep{2014AJ....147..136R} and $0.044$ \citep{2015MNRAS.454.2863H} while $\psi$~Phe has ${E(B-V) = 0.026}$ \citep{2015MNRAS.454.2863H}.
In both cases the estimated reddenings are so small as to make no significant difference to our age estimates.
For $\beta$~Ara we were unable to find any reddening estimates in the literature.
Based on these investigations and the fact that most stars are inside the Local Bubble we have decided not to apply any reddening corrections to the $V$ magnitudes used in this study.

\begin{table*}
  \centering
  \caption{Stellar parameters for the \textit{Gaia} Benchmark stars adopted for the derivation of ages from Bayesian isochrone fitting.
  All metallicities are taken from Table 3 in \citet{2014A&A...564A.133J}, and all values of $T_{\mathrm{eff}}$ and $\log g$ are taken from Table 10 in \citet{2015A&A...582A..49H} unless the value is marked by an asterisk.
  The parallaxes come from the revised \textit{Hipparcos} catalogue \citep{2007ASSL..350.....V} for most stars, from \citet{1999A&A...341..121S} for $\alpha$~Cen~A \& B, and from \citet{2014ApJ...792..110V} for the two stars HD~84937 and HD~140283.
  $V$ band magnitudes come from the SIMBAD database.
  The stars are ordered by their spectral classification following \citet{2015A&A...582A..49H}.}
  \label{tbl:obs_data}
  \begin{threeparttable}
  \begin{tabular}{lrrrrrrrrrrrr}
    \hline
    Name  & HD  & HIP & [Fe/H]   & $\sigma$([Fe/H]) & $T_{\mathrm{eff}}$ [K] & $\sigma(T_{\mathrm{eff}})$ &
    $\log g$&$\sigma(\log g)$&$\varpi$ [mas] & $\sigma(\varpi)$ & $V$mag & $\sigma$($V$mag)\\ \hline
    Procyon        & $61421$  & 37279 & $0.01$  & $0.08$     & $6554$ & $84$        &
    $4.00$ & $0.02$      & $284.52$ & $1.27$     & $0.37$  & $0.02$   \\
    HD 84937       & 84937  & 48152 & $-2.03$ & $0.08$     & $6356$ & $97$        &
    $4.06$ & $0.04$      & $12.24$  & $0.20$     & $8.32$  & $0.02$   \\
    HD 49933       & $49933$  & 32851 & $-0.41$ & $0.08$     & $6635$ & $91$        &
    $4.20$ & $0.03$      & $33.68$  & $0.42$     & $5.78$  & $0.02$   \\
    $\delta$ Eri   & 23249  & 17378 & $0.06$  & $0.05$     & $4954$ & $30$        &
    $3.76$ & $0.02$      & $110.62$ & $0.22$     & $3.54$  & $0.02$   \\
    HD 140283      & 140283 & 76976 & $-2.36$ & $0.10$     & $5692$\tnote{*} & $102$\tnote{*}       &
    $3.58$ & $0.11$      & $17.18$  & $0.26$     & $7.21$  & $0.02$   \\
    $\epsilon$ For & 18907  & 14086 & $-0.60$ & $0.10$     & $5123$ & $78$        &
    $4.07$\tnote{*} & $0.30$\tnote{*}      & $31.05$  & $0.36$     & $5.85$  & $0.02$   \\
    $\eta$ Boo     & 121370 & 67927 & $0.32$  & $0.08$     & $6099$ & $28$        &
    $3.79$ & $0.02$      & $87.77$  & $1.24$     & $2.68$  & $0.01$   \\
    $\beta$ Hyi    & 2151   & 2021 & $-0.04$ & $0.06$     & $5873$ & $45$        &
    $3.98$ & $0.02$      & $134.07$ & $0.11$     & $2.79$  & $0.02$   \\
    $\alpha$ Cen A & 128620 & 71683 & $0.26$  & $0.08$     & $5792$ & $16$        &
    $4.31$ & $0.01$      & $747.10$ & $1.20$     & $0.01$  & $0.02$   \\
    HD 22879       & 22879  & 17147 & $-0.86$ & $0.05$     & $5868$ & $89$        &
    $4.27$ & $0.04$      & $39.13$  & $0.57$     & $6.67$  & $0.02$   \\
    $\mu$ Cas      & 6582   & 5336 & $-0.81$ & $0.03$     & $5308$ & $29$        &
    $4.51$\tnote{*} & $0.04$\tnote{*}      & $132.40$ & $0.82$     & $5.17$  & $0.02$   \\
    $\tau$ Cet     & 10700  & 8102 & $-0.49$ & $0.03$     & $5414$ & $21$        &
    $4.58$\tnote{*} & $0.02$\tnote{*}      & $273.96$ & $0.17$     & $3.50$  & $0.02$   \\
    $\alpha$ Cen B & 128621 & 71681 & $0.22$  & $0.10$     & $5231$ & $20$        &
    $4.53$ & $0.03$      & $747.10$ & $1.20$     & $1.33$  & $0.02$   \\
    18 Sco         & 146233 & 79672 & $0.03$  & $0.03$     & $5810$ & $80$        &
    $4.44$ & $0.03$      & $71.93$  & $0.37$     & $5.50$  & $0.02$   \\
    $\mu$ Ara      & 160691 & 86796 & $0.35$  & $0.13$     & $5783$\tnote{*} & $46$\tnote{*}        &
    $4.30$ & $0.03$      & $64.48$  & $0.31$     & $5.15$  & $0.02$   \\
    $\beta$ Vir    & 102870 & 57757 & $0.24$  & $0.07$     & $6083$ & $41$        &
    $4.10$ & $0.02$      & $91.50$  & $0.22$     & $3.60$  & $0.02$   \\
    Arcturus       & 124897 & 69673 & $-0.52$ & $0.08$     & $4286$ & $35$        &
    $1.60$\tnote{*} & $0.20$\tnote{*}      & $88.83$  & $0.53$     & $-0.05$ & $0.02$   \\
    HD 122563      & 122563 & 68594 & $-2.64$ & $0.22$     & $4587$ & $60$        &
    $1.61$ & $0.07$      & $4.22$   & $0.35$     & $6.19$  & $0.02$   \\
    $\mu$ Leo      & 85503  & 48455 & $0.25$  & $0.15$     & $4474$ & $60$        &
    $2.51$ & $0.11$      & $26.27$  & $0.16$     & $3.88$  & $0.02$   \\
    $\beta$ Gem    & 62509  & 37826 & $0.13$  & $0.16$     & $4858$ & $60$        &
    $2.90$ & $0.08$      & $96.52$  & $0.24$     & $1.14$  & $0.02$   \\
    $\epsilon$ Vir & 113226 & 63608 & $0.15$  & $0.16$     & $4983$ & $61$        &
    $2.77$ & $0.02$      & $29.75$  & $0.14$     & $2.79$  & $0.02$   \\
    $\xi$ Hya      & 100407 & 56343 & $0.16$  & $0.20$     & $5044$ & $40$        &
    $2.87$ & $0.02$      & $25.14$  & $0.16$     & $3.54$  & $0.02$   \\
    HD 107328      & 107328 & 60172 & $-0.33$ & $0.16$     & $4496$ & $59$        &
    $2.09$ & $0.13$      & $10.60$  & $0.25$     & $4.96$  & $0.02$   \\
    HD 220009      & 220009 & 115227 & $-0.74$ & $0.13$     & $4402$\tnote{*} & $111$\tnote{*}       &
    $1.95$\tnote{*} & $0.34$\tnote{*}      & $7.55$   & $0.40$     & $5.07$  & $0.01$   \\
    $\alpha$ Tau   & 29139  & 21421 & $-0.37$ & $0.17$     & $3927$ & $40$        &
    $1.11$ & $0.19$      & $48.92$  & $0.77$     & $0.86$  & $0.02$   \\
    $\alpha$ Cet   & 18884  & 14135 & $-0.45$ & $0.47$     & $3796$ & $65$        &
    $0.68$ & $0.23$      & $13.10$  & $0.44$     & $2.53$  & $0.02$   \\
    $\beta$ Ara    & 157244 & 85258 & $-0.05$ & $0.39$     & $4197$ & $50$        &
    $1.05$ & $0.15$      & $4.54$   & $0.61$     & $2.85$  & $0.02$   \\
    $\gamma$ Sge   & 189319 & 98337 & $-0.17$ & $0.39$     & $3807$ & $49$        &
    $1.05$ & $0.32$      & $12.61$  & $0.18$     & $3.47$  & $0.02$   \\
    $\psi$ Phe     & 11695  & 8837 & $-1.24$ & $0.39$     & $3472$ & $92$        &
    $0.51$ & $0.18$      & $9.54$   & $0.20$     & $4.41$  & $0.02$   \\
    $\epsilon$ Eri & 22049  & 16537 & $-0.09$ & $0.06$     & $5076$ & $30$        &
    $4.61$ & $0.03$      & $310.95$ & $0.16$     & $3.73$  & $0.02$   \\
    Gmb 1830       & 103095 & 57939 & $-1.46$ & $0.39$     & $5087$\tnote{*} & $100$\tnote{*}       &
    $4.60$ & $0.03$      & $109.98$ & $0.41$     & $6.45$  & $0.02$   \\
    61 Cyg A       & 201091 & 104214 & $-0.33$ & $0.38$     & $4374$ & $22$        &
    $4.63$ & $0.04$      & $286.83$ & $6.77$     & $5.21$  & $0.02$   \\
    61 Cyg B       & 201092 & 104217 & $-0.38$ & $0.03$     & $4044$ & $32$        &
    $4.67$ & $0.04$      & $285.89$ & $0.55$     & $6.03$  & $0.02$   \\ \hline
  \end{tabular}
  \begin{tablenotes}
  \small
  \item[*] Value not taken from Table 10 in \citet{2015A&A...582A..49H}, see Appendix~\ref{sec:appA} for the sources for each individual star.
  \end{tablenotes}
  \end{threeparttable}
\end{table*}

\subsection{Summary of the fitting method} \label{sec:fit_summary}
The method used to determine stellar ages in this work is identical in its formalism to the Bayesian fitting algorithm developed by \citet{2018arXiv180408321H} for the purpose of calculating the two-dimensional $\mathcal{G}$~function, $\mathcal{G}(\tau, \zeta|\mathbf{x})$.
This function is the joint relative likelihood of the age, $\tau$, and metallicity, $\zeta$, given a set of observed stellar parameters $\mathbf{x}$, and it is a generalisation of the one-dimensional case, $\mathcal{G}(\tau|\mathbf{x})$, introduced by \citet{2005A&A...436..127J}.
In short, the $\mathcal{G}$~function is calculated by fitting stellar models (isochrones in this case) to a set of observables, $\mathbf{x}$.
For the present purpose, the stellar models are described by their initial mass, $m$, age, $\tau$, metallicity, $\zeta$, and distance modulus, $\mu$, and the observables are one of the two sets described in Section~\ref{sec:obs_data} ($\log g$ or the $V$ magnitude in addition to $T_{\mathrm{eff}}$ and [Fe/H]).
The addition of $\mu$ to the model parameters allows the stellar parallax to be included directly in the likelihood calculation.
The likelihood of each model in a grid is calculated based on the observables, and $\mathcal{G}(\tau, \zeta|\mathbf{x})$ is obtained by marginalising over $m$ and $\mu$ with suitable prior densities.
For $m$ we use a Salpeter initial mass function as prior, and we use a flat prior on $\mu$.
For a complete description of the formalism, see Appendix~A in \citet{2018arXiv180408321H}.

In this work we are only interested in the age dimension, and the observed metallicity is included in the likelihood of each model.
Therefore, after calculating what is formally the same as the two-dimensional $\mathcal{G}$~function -- but different in the sense that we have already constrained the metallicity -- we marginalise over the metallicity dimension with a flat prior.
This yields the one-dimensional function $\mathcal{G}(\tau|\mathbf{x})$ from which the age of the star can be estimated.

\begin{figure}
  \center
  \includegraphics[width=\columnwidth]{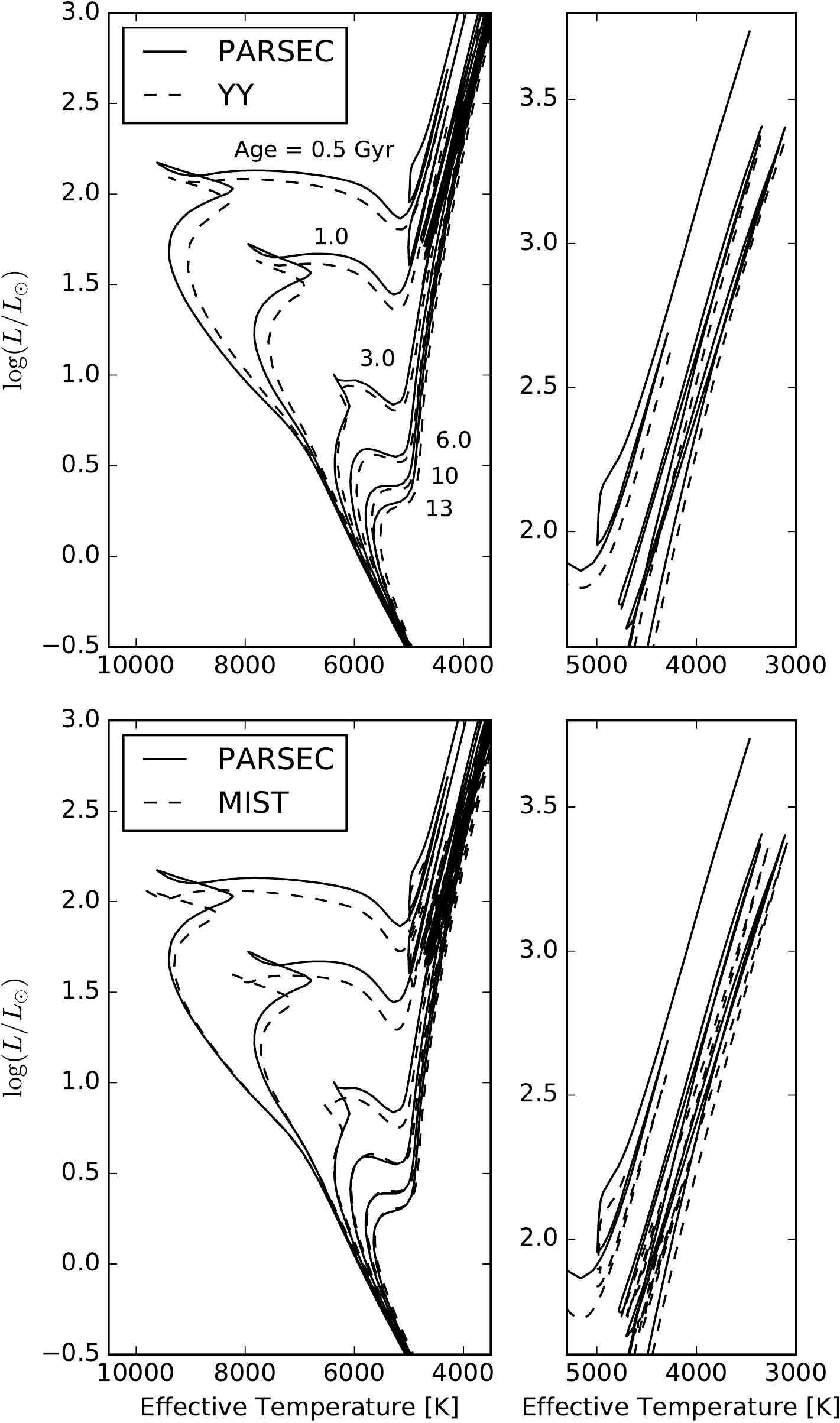}
  \caption{HR~diagrams comparing the PARSEC isochrones with the $Y^2$ isochrones (top row) and with the MIST isochrones (bottom row) at solar metallicity.
  In the left-hand column isochrones of ages 0.5, 1, 3, 6, 10, and 13~Gyr are shown, and in the right-hand column  isochrones of ages 0.5, 3, and 10~Gyr are shown for the giant branch.}
  \label{fig:iso_plot}
\end{figure}

\begin{table}
  \centering
  \caption{Parameter ranges and resolution of the three isochrone grids given in the format min(step)max.}
  \label{tbl:iso}
  \begin{tabular}{lll}
    \hline
    Isochrones & Ages {[}Gyr{]} & [Fe/H] \\ \hline
    $Y^2$         & $0.1(0.1)15$    & $-2.5(0.05)0.5$         \\
    PARSEC     & $0.1(0.1)13.5$  & $-2.5(0.05)0.5$         \\
    MIST       & $0.1(0.1)16$    & $-3.0(0.05)0.5$         \\ \hline
    \end{tabular}
\end{table}

\subsection{Isochrones} \label{sec:isochrones}
Since the isochrone-based ages depend on the adopted stellar models, we have chosen to calculate the $\mathcal{G}$~functions for each star using three different sets of isochrones.
These three sets are Yonsei-Yale\footnote{\url{http://www.astro.yale.edu/demarque/yyiso.html}} ($Y^2$, \citealt{2004ApJS..155..667D}), PARSEC\footnote{\url{http://stev.oapd.inaf.it/cmd}} \citep{2012MNRAS.427..127B, 2014MNRAS.444.2525C, 2015MNRAS.452.1068C, 2014MNRAS.445.4287T}, and MIST\footnote{\url{http://waps.cfa.harvard.edu/MIST/interp_isos.html}} \citep{2016ApJ...823..102C, 2016ApJS..222....8D} which are based on the MESA code \citep{2011ApJS..192....3P, 2013ApJS..208....4P, 2015ApJS..220...15P}.
These three sets of isochrones are all based on models including microscopic diffusion (although only of helium in $Y^2$) and convective core overshoot.
In this work we are not interested in exploring the effects of turning such physical processes off, we only want a sense of the impact on the ages of the more subtle changes in e.g. the solar abundance scale or the convective overshoot efficiency which are parameters that are not yet fully understood.
In this context it can be mentioned that the PARSEC isochrones are based on the solar composition of \citet{1998SSRv...85..161G} complemented for some elements by \citet{2011SoPh..268..255C} whereas the MIST isochrones are based on the solar composition of \citet{2009ARA&A..47..481A}.
They also differ in the adopted helium-to-metal enrichment ratio which changes the composition of models with metallicities different from solar. The values of $\Delta Y / \Delta Z$ are 1.50, 1.78, and 2.00 for MIST, PARSEC, and YY, respectively.
The combination of different solar compositions and helium-to-metal enrichment ratios means that models with the same value of [Fe/H] have different compositions (X, Y, and Z) in the three sets of isochrones.

For each of the three sets we have created a grid of isochrones with the same resolution of 0.1~Gyr in age and 0.05~dex in metallicity.
Due to the very precise observables, the default mass resolutions turned out to be too coarse for some of the stars on the main-sequence and subgiant branch.
Therefore, we have interpolated the isochrones (at fixed age and metallicity) onto a denser grid of masses for models below the giant branch.
The grids differ slightly in the range of ages and metallicities included: these values are summarised in \autoref{tbl:iso}.
A more significant difference is found in the evolutionary stages included in the grids.
The $Y^2$ grid does not include models beyond the tip of the red giant branch (RGB) while the PARSEC and MIST grids include the more advanced stages of the red clump (RC) and asymptotic giant
branch (AGB).
This has important consequences for the age determination of giants as will be discussed in Section~\ref{sec:g_functions}.

The three different isochrone grids are compared in \autoref{fig:iso_plot} for models with solar metallicity and a number of different ages.
On the main sequence and the turn-off of the old isochrones, the $Y^2$ isochrones are generally cooler than the other two sets which will lead to slightly lower ages for most stars.
However, after the turn-off of the young isochrones, the PARSEC models differ from the other two by having a more luminous subgiant branch at a given age.
Since these younger models are more massive and have convective cores, this is most likely due to differences in the treatment of convective core overshooting.
Changing the overshooting efficiency changes the convective core mass; this in turn affects the main-sequence lifetime and hence the position of the turnoff in the HR~diagram
On the giant branch the PARSEC isochrones are hotter than the other two sets which will lead to slightly higher ages for giants.

The comparison in \autoref{fig:iso_plot} is only for isochrones at solar metallicity, and at different metallicities they do not show exactly the same trends.
For example, at a metallicity of $-2$~dex the MIST isochrones are the ones that stand out as being slightly hotter overall at a given age.
This means that the MIST isochrones will predict the highest ages for metal-poor stars.
Other than that, the comparison does not change much with metallicity.

\subsection{Special cases} \label{sec:special_cases}
In some cases, the method we have outlined so far falls short either due to inadequacies in the isochrones or in the fitting method.
Here we address these cases and describe additional fits we have carried out to investigate their influence on the final results.

\subsubsection{$\beta$~Ara} \label{sec:special_beta_ara}
One of the benchmark stars, $\beta$~Ara, falls outside of our grids due to its high mass ($8.21~M_{\odot}$ according to \citet{2015A&A...582A..49H}).
For this star we created a small MIST grid of younger isochrones with the same metallicities as the main grid, but with ages in the range 0.01--0.50~Gyr in steps of 0.01~Gyr.
With this grid we are able to fit the star, but the results based on the three main grids are not reliable.
Therefore, for the final age of this star, we only consider the results obtained using this young MIST grid.

\begin{figure*}
  \center
  \includegraphics[width=0.88\textwidth]{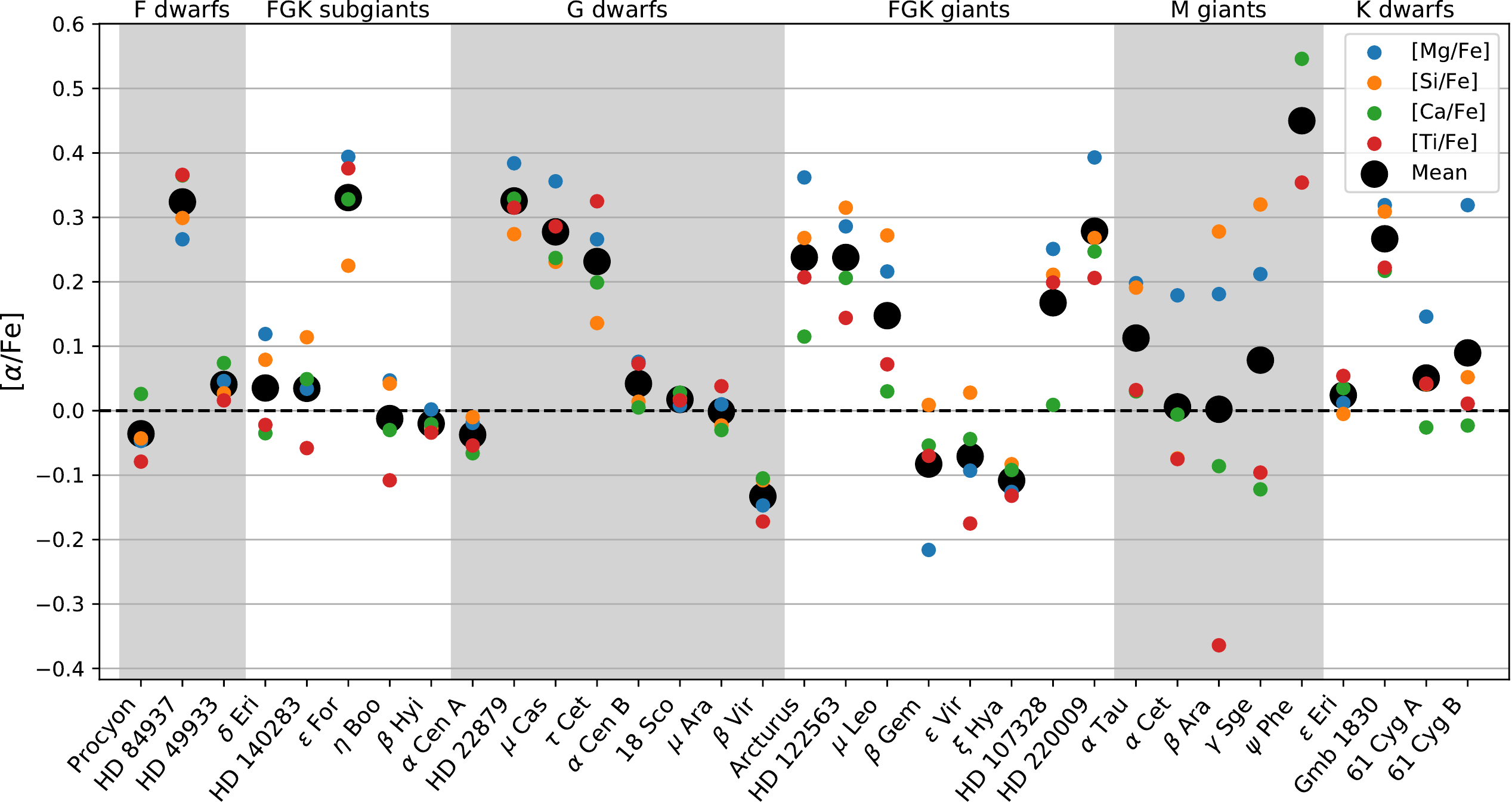}
  \caption{Abundances for the four $\alpha$-elements magnesium, silicon, calcium, and titanium for the benchmark stars from \citet{2015A&A...582A..81J}.
  The small circles show abundances of the individual elements, and the large black circles are the mean values of the four elements as indicated in the legend.
  Only two abundances are available for $\psi$~Phe.}
  \label{fig:alpha_abundances}
\end{figure*}

\subsubsection{$\alpha$-enhanced stars}
The isochrone grids presented above are all based on solar-scaled abundances; however, some of the benchmark stars are enhanced in $\alpha$-elements relative to the Sun.
A change in the $\alpha$-abundances also changes the stellar parameters including the age.
Therefore, we have created additional $Y^2$ grids with increased [$\alpha$/Fe] (0.2 and 0.4~dex) to investigate the significance of this effect.

We estimate [$\alpha$/Fe] for each of the benchmark stars by considering the abundances determined by \citet{2015A&A...582A..81J} for the four $\alpha$-elements magnesium, silicon, calcium, and titanium.
They give values of [X/H] for each of the elements, and by subtracting [Fe/H] the values of [X/Fe] shown in \autoref{fig:alpha_abundances} are obtained.
We use the mean values of the four elements as estimates of [$\alpha$/Fe] to decide which of the stars should be fitted to the $\alpha$-enhanced $Y^2$ grids.
Four stars have a mean value above 0.3~dex; these are fitted to the isochrones with [$\alpha$/Fe] $=$ 0.4~dex.
Another eight stars have mean values between 0.1 and 0.3~dex (excluding $\alpha$~Tau which is right on the limit); these are fitted to the isochrones with [$\alpha$/Fe] $=$ 0.2~dex.
The results of these fits are included in the individual discussions in Appendix~\ref{sec:appA}.
For most of these stars the inclusion of $\alpha$-enhancement leads to a negligible change in age, but for some of the oldest ones it turns out to be important in order to keep the observed parameters within the limits of the model grids (this is discussed in Section~\ref{sec:old_mpoor}). 

\subsubsection{Fitting to the current surface metallicity}
A subtle detail of the fitting procedure which has not yet been addressed is the fact that we compare the observed metallicity to the initial metallicities of the stellar models when calculating the model likelihoods.
In reality, the surface metallicity changes with evolution and if we were to fit to the current surface metallicity, the age would change in some cases.
In fact, fitting to the initial surface metallicity instead of the current one has been shown to lead to overestimation of the age by up to 20 per cent at the turn-off \citep{2017ApJ...840...99D}.
The reason we fit to the initial metallicity anyway is that the surface metallicities of the models are missing from the $Y^2$ and PARSEC isochrones.
In the MIST isochrones, however, both the initial and current surface metallicity are available for each model.
Therefore, we have repeated the MIST fits using the current surface metallicities of the models instead.
This has been done for all stars, and we include the results in the individual discussions of the stars in Appendix~\ref{sec:appA}.
Overall, the impact of this change is minor and completely negligible for almost all of the stars.
The only exception is the turn-off star HD~49933 (F~dwarf) for which the age is decreased by about 1--2~Gyr which is more than the statistical uncertainty.
We believe the effect is significant for this star because the impact is greater at higher temperatures as shown by \citet{2017ApJ...840...99D}, and HD~49933 is the hottest of the turn-off stars in the benchmark sample.

\begin{figure*}
  \center
  \includegraphics[width=\textwidth]{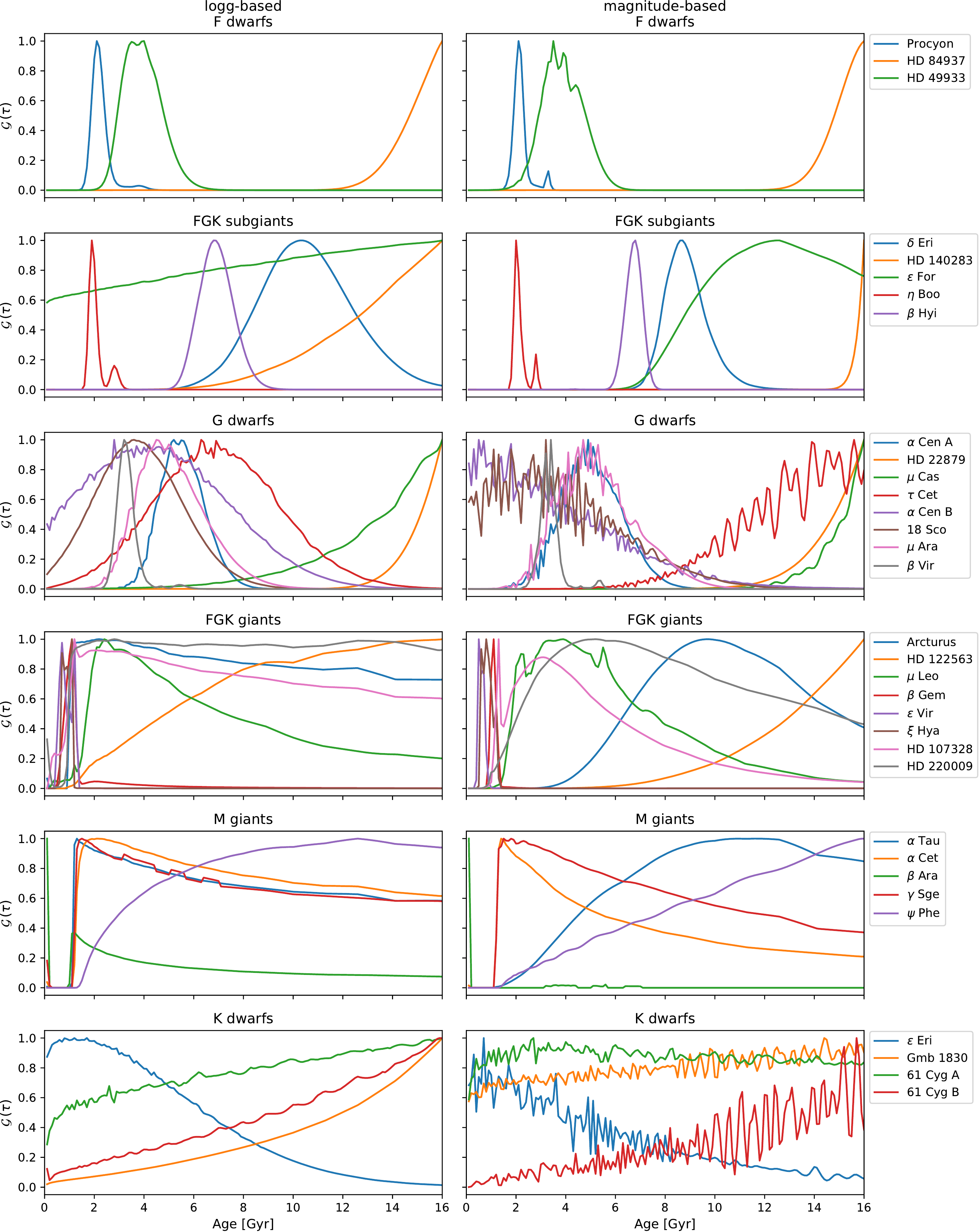}
  \caption{$\mathcal{G}$~functions for the \textit{Gaia} benchmark stars sorted according to their spectral classification.
  These are all based on the MIST isochrones and the left-hand column is $\log g$-based and the right-hand column is magnitude-based.
  For each star, the corresponding $\mathcal{G}$~functions are given the same colour in both columns.}
  \label{fig:gfuncs_MIST}
\end{figure*}

\begin{figure}
  \center
  \includegraphics[width=0.9\columnwidth]{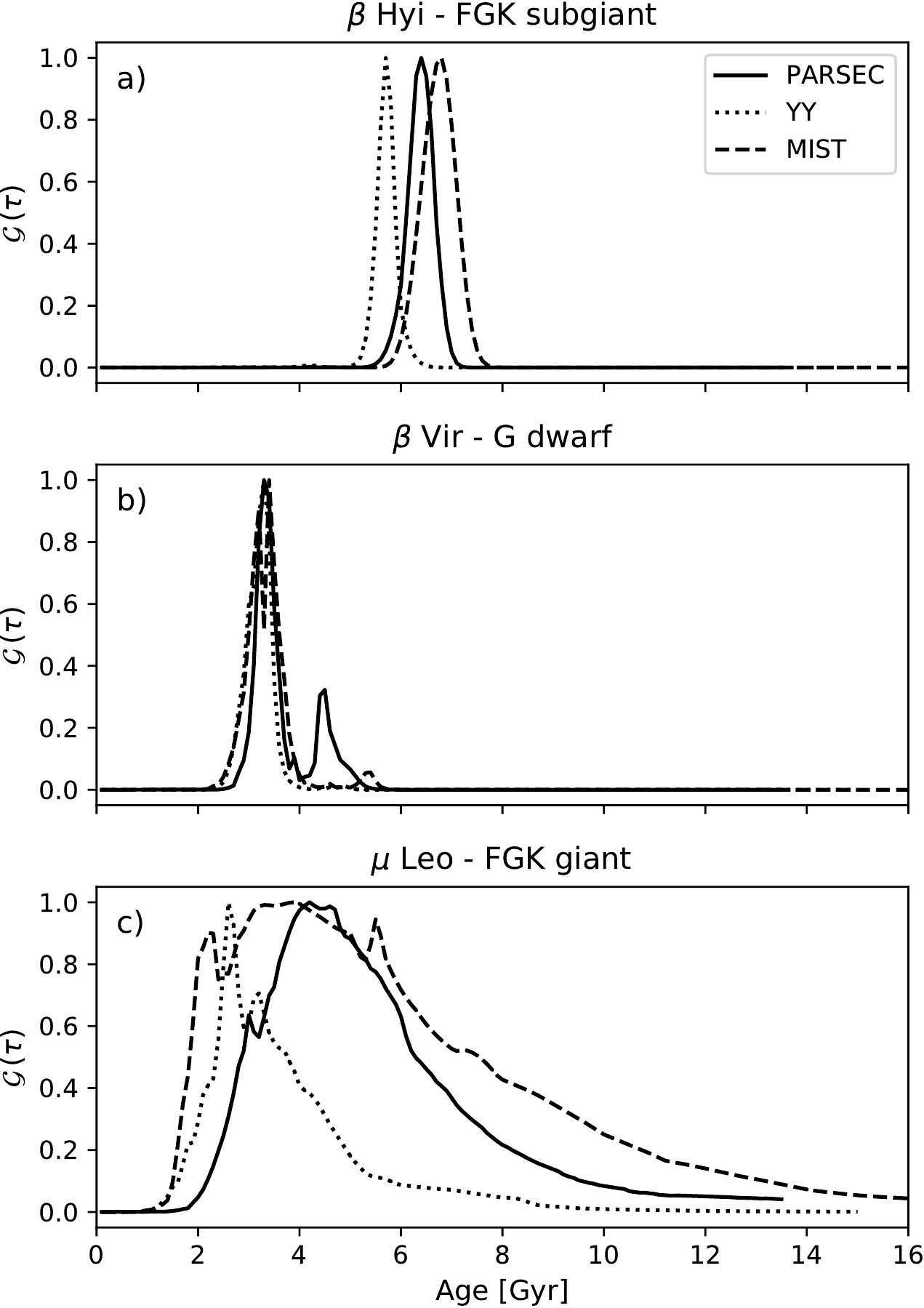}
  \caption{Magnitude-based $\mathcal{G}$~functions for: a) $\beta$~Hyi, b) $\beta$~Vir, and c) $\mu$~Leo.
  The different line styles indicate results based on the three different isochrone grids as given in the legend.}
  \label{fig:gfuncs_subset}
\end{figure}

\subsection{G-functions across the HR~diagram} \label{sec:g_functions}
As mentioned previously, the output of our Bayesian fitting algorithm is the $\mathcal{G}$~function, $\mathcal{G}(\tau|\mathbf{x})$.
Assuming a flat prior on the age, the $\mathcal{G}$~function is proportional to the posterior density of the age.
It then describes the relative probability of different stellar ages given a set of observations and isochrones, and we normalise it to a maximum value of 1.
With the combination of three different grids of isochrones and two sets of observables, we have calculated six different $\mathcal{G}$~functions for each star (not including the special cases discussed in Section~\ref{sec:special_cases}).
In \autoref{fig:gfuncs_MIST} two out of the six are shown, namely the ones calculated using the MIST isochrones based on either $\log g$ (left-hand column) or the magnitude (right-hand column).
This gives a sense of the shapes of the $\mathcal{G}$~functions for different classes of stars.

Not surprisingly, the most well-constrained ages are found among the F~dwarfs and FGK~subgiants which are located in the regions of the HR~diagram where the isochrones are well separated.
For the $\mathcal{G}$~functions with modes below 12~Gyr, the shapes are approximately Gaussian which allows us to obtain an age estimate including reliable uncertainties.
In the very best cases (Procyon, $\eta$~Boo, and $\beta$~Hyi), the relative magnitude-based age uncertainties are about 5 per cent owing to the precise observables and the favourable positions in the HR~diagram.
However, for these precise age estimates, the choice of isochrones still limits the accuracy.
As an example, the magnitude-based $\mathcal{G}$~functions of $\beta$~Hyi are shown for the three different grids of isochrones in \autoref{fig:gfuncs_subset}a.
In this case the modes of the distributions vary by about 1~Gyr which is significant compared to the statistical uncertainty of about 0.25~Gyr.

The $\mathcal{G}$~functions are wider for older stars and some of them have the mode at the old edge of the grid.
These old stars are also among the most metal-poor ones in the sample (e.g. HD~84937 and 140283) and we discuss their ages in Section~\ref{sec:old_mpoor}.
The two oldest subgiants have the least precise values of $\log g$ which leads to their more extended $\mathcal{G}$~functions in the left-hand panel.
The one that is almost flat is for $\epsilon$ For, but we believe this is due to the adopted value of $\log g$ being too high (see the discussion of this star in Appendix~\ref{sec:appA}).

Moving down in luminosity to the G~dwarfs, the function shapes show larger variations since this category hosts both stars on the main-sequence and stars close to the turn-off.
The magnitude-based results give Gaussian-like $\mathcal{G}$~functions for only three of the stars ($\alpha$~Cen~A, $\mu$~Ara, and $\beta$~Vir).
These are the ones that have moved off the main sequence and towards the turn-off.
As shown in \autoref{fig:gfuncs_subset}b, $\beta$~Vir shows a bimodal $\mathcal{G}$~function when using the PARSEC isochrones.
This difference in shape is due to the differences between the isochrones around the turn-off which were pointed out in Section~\ref{sec:isochrones}.
The rest of the G~dwarfs are either at the old edge of the isochrone grids or further down the main sequence where the isochrones converge.
Due to a decrease in the number of models at lower masses, the $\mathcal{G}$~functions become increasingly spiky for the low-mass main sequence stars.
For the $\log g$-based results, a larger fraction of the $\mathcal{G}$~functions approximate Gaussians.
This is a reflection of the fact that on the main sequence the isochrones are slightly better separated by
$\log g$ compared to the luminosity.
In combination with the very precise stellar parameters available for this sample, this makes the $\log g$-based ages more well-defined for some of these stars (e.g. $\tau$~Cet, $\alpha$~Cen~B, and
18~Sco).
Two of the G~dwarfs (HD~22879 and $\mu$~Cas) have modes at the highest age in the grid.
Just like the oldest stars among the F~dwarfs and FGK~subgiants, these two stars are the most metal-poor of the G~dwarfs.

For the giants, the $\mathcal{G}$~functions generally split into two categories: those with a well-defined peak at $\lesssim 2$~Gyr, and those with an extended distribution at higher ages.
This reflects a separation of the isochrones on the giant branch where, at a given metallicity, the younger isochrones ($\lesssim 2$~Gyr) show no overlap with the older ones.
For slightly older isochrones (e.g. 3~Gyr), the RGB passes through the RC of the older isochrones (see e.g. \autoref{fig:iso_plot}) which greatly increases the uncertainty in the age.
At the same time the older isochrones are more closely spaced just like in the turn-off region.
The spacing is smaller in $\log g$ than in magnitude which leads to the $\log g$-based $\mathcal{G}$~functions being almost completely flat for many of the giants.
The $\mathcal{G}$~functions of the M~giants are generally less informative than those of the FGK~giants due to larger uncertainties on [Fe/H] and $\log g$.

In \autoref{fig:gfuncs_subset}c the effect of the overlapping evolutionary stages is shown for $\mu$~Leo.
Since our grid of $Y^2$ isochrones does not contain evolutionary stages beyond the RGB, it gives a relatively well-constrained age estimate.
However, this star is located right on top of the RC in the HR~diagram, so when using the other two grids where the RC phase is included, higher ages have a significant probability as well.
Thus, without prior knowledge about the evolutionary stage of the star, the age is not very well constrained.

Finally, we have the K~dwarfs.
As expected, these stars are all too far down the main sequence for their stellar parameters to reveal any useful age information.
This results in $\mathcal{G}$~functions which are either almost completely flat or peak at one end of the age interval with a long tail towards the other end.
Due to the convergence of the isochrones on the main sequence, slight changes in the stellar parameters can significantly change the shape of these $\mathcal{G}$~functions; therefore, these functions do not put any limits on the stellar age even when they fall off towards one end of the age interval.

There are different ways to obtain a single age estimate based on the $\mathcal{G}$~function.
In the figures accompanying the discussions of individual stars in Appendix~\ref{sec:appA}, age estimates are shown based on the mode of the G~function following \citet{2005A&A...436..127J}.
We also follow their method for determining the 68 per cent confidence interval as the region within which the $\mathcal{G}$~function is larger than 0.6.
Among other desirable properties, this choice ensures that the age estimate is within the interval defined by the lower and upper confidence limits (see \citealt[section~3.6]{2005A&A...436..127J} for further discussion).
It also means that either or both of the confidence limits can be undefined when the $\mathcal{G}$~function is above 0.6 at the edges of the isochrone grids.
In these cases no well-defined age can be assigned to the star.
Instead of publishing the ages based on our own choice of statistics, we make available all six $\mathcal{G}$~functions for each star in the online material.

\begin{figure}
  \center
  \includegraphics[width=\columnwidth]{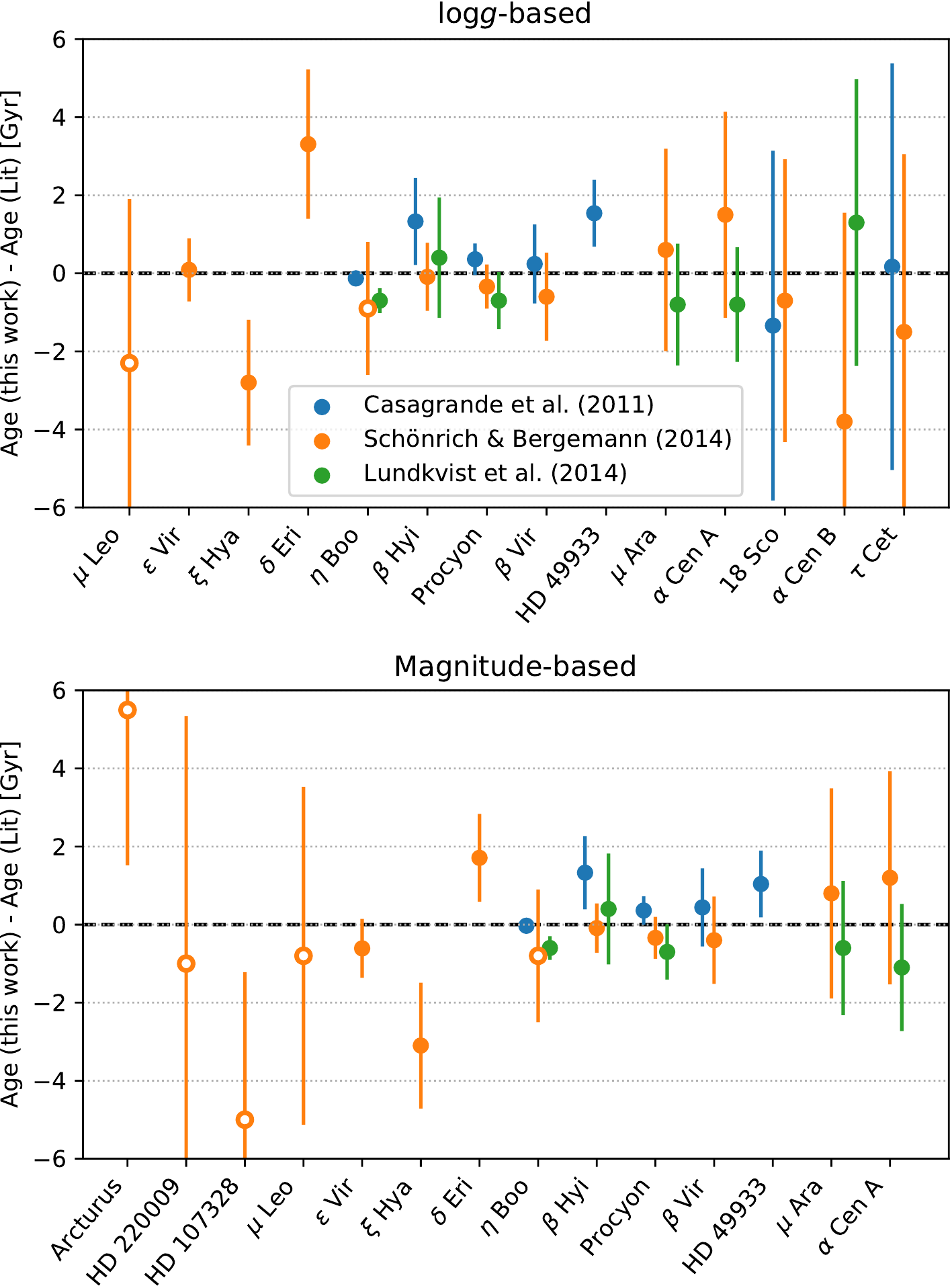}
  \caption{Differences between ages derived in this work (based on MIST isochrones) and ages in the literature by \citet{2014MNRAS.443..698S, 2011A&A...530A.138C, 2014A&A...566A..82L}.
  The stars are shown in order of increasing $\log g$ from left to right.
  In the upper panel the literature is compared with $\log g$-based ages and in the lower panel with magnitude-based ages, but the literature ages are the same in both panels.
  In both cases only the stars for which the age derived in this work is well-defined are shown which means that different stars are present in the two panels. The open symbols indicate stars for which \citet{2014MNRAS.443..698S} included only photometric constraints in their fit (i.e. no spectra).}
  \label{fig:lit_comp}
\end{figure}

\subsection{Comparisons with the literature} \label{sec:age_lit_comp}
\autoref{fig:lit_comp} shows a comparison between ages determined in this work from Bayesian isochrone fitting to MIST models and a few different literature values.
The comparison is only made for the stars for which the estimate from isochrone fitting has a well-defined 68 per cent confidence interval which means that the $\mathcal{G}$~function falls below a value of 0.6 on both sides of the mode.
This means that different stars are shown in the $\log g$- and magnitude-based comparisons; for example, $\alpha$~Cen~B has a well-defined confidence interval in the fit to $\log g$ but not the magnitude (see \autoref{fig:gfuncs_MIST}).

First of all we compare the isochrone-based ages with those determined by \citet{2014MNRAS.443..698S} since they determined ages for most of the benchmark stars, but with a different fitting method which was briefly summarised in Section~\ref{sec:lit_fitting}.
For this comparison, the stars for which only photometry was used in the fit by \citeauthor{2014MNRAS.443..698S} are marked with open symbols.
These stars are mostly older ($>2$~Gyr) giants for which the isochrone-based ages are uncertain even when spectroscopic information is included, and this is where the largest age differences are seen.
The subgiant $\eta$~Boo was also only fitted to photometry in their analysis, but the age estimates still agree to within 1~Gyr although their uncertainty is about 60 per cent compared to 5 per cent on our estimate.
For the stars which they fitted with spectra there is generally good agreement with increasing differences for the more uncertain ages of the dwarf stars.
For $\xi$~Hya and $\delta$~Eri the ages differ by more than 1$\sigma$.
\citeauthor{2014MNRAS.443..698S} state that their metallicity fit is questionable for $\xi$~Hya and that they disregarded a bad spectral fit for $\delta$~Eri which indicates that these stars were difficult to fit to their spectral data.

We also compare with ages of the GCS stars by \citet{2011A&A...530A.138C}, which is our second largest source of literature ages and based on a Bayesian algorithm similar to the one used in this work; and with \citet{2014A&A...566A..82L} which is our largest source of ages with asteroseismic constraints.
The ages determined in this work agree well (within 2~Gyr) with the ages from both of these studies, which is to be expected since most of them are subgiants and dwarfs near the turn-off.

\begin{table}
  \centering
  \caption{Benchmark ages, i.e. ages for the \textit{Gaia} benchmark stars based on the combined age information collected and derived in this study.
  Each age is given as a range with a lower and upper limit as well as a rank of A, B, or C (see the text for an explanation).
  For each star the discussion leading to these age ranges is given in Appendix \ref{sec:appA}.}
  \label{tbl:benchmark_ages}
  \begin{tabular}{lrrrrr}
    \hline
    Name  & HD  & HIP & \multicolumn{2}{c}{Age [Gyr]} & Rank \\ 
    \cline{4-5}
                &        &        & Min        & Max                       &  \\\hline
    Procyon        & 61421  & 37279 & $1.5$  & $2.5$     & A \\
    HD 84937       & 84937  & 48152 & $11.0$ & $13.5$     & A   \\
    HD 49933       & 49933  & 32851 & $2.0$ & $4.0$     & A   \\
    $\delta$ Eri   & 23249  & 17378 & $6.0$  & $9.0$     & A   \\
    HD 140283      & 140283 & 76976 & $12.0$ & ---     & B   \\
    $\epsilon$ For & 18907  & 14086 & $4.0$ & $12.0$     & B   \\
    $\eta$ Boo     & 121370 & 67927 & $2.0$  & $3.0$     & A   \\
    $\beta$ Hyi    & 2151   & 2021 & $5.0$ & $7.0$     & A   \\
    $\alpha$ Cen A & 128620 & 71683 & $4.0$  & $7.0$     & A   \\
    HD 22879       & 22879  & 17147 & $8.0$ & ---     & B   \\
    $\mu$ Cas      & 6582   & 5336 & $3.0$ & ---     & C   \\
    $\tau$ Cet     & 10700  & 8102 & $4.0$ & $10.0$     & B   \\
    $\alpha$ Cen B & 128621 & 71681 & $4.0$  & $7.0$     & A   \\
    18 Sco         & 146233 & 79672 & $3.0$  & $5.0$     & A   \\
    $\mu$ Ara      & 160691 & 86796 & $4.0$  & $8.0$     & A   \\
    $\beta$ Vir    & 102870 & 57757 & $2.0$  & $4.0$     & A   \\
    Arcturus       & 124897 & 69673 & $4.0$ & $10.0$     & B   \\
    HD 122563      & 122563 & 68594 & --- & ---     & C   \\
    $\mu$ Leo      & 85503  & 48455 & $2.0$  & $7.0$     & B   \\
    $\beta$ Gem    & 62509  & 37826 & $0.8$  & $1.5$     & A   \\
    $\epsilon$ Vir & 113226 & 63608 & $0.4$  & $1.2$     & A   \\
    $\xi$ Hya      & 100407 & 56343 & $0.5$  & $1.0$     & A   \\
    HD 107328      & 107328 & 60172 & $1.0$ & $10.0$     & B   \\
    HD 220009      & 220009 & 115227 & $2.0$ & ---     & C   \\
    $\alpha$ Tau   & 29139  & 21421 & $2.0$ & ---     & C   \\
    $\alpha$ Cet   & 18884  & 14135 & $1.0$ & $10.0$     & C   \\
    $\beta$ Ara    & 157244 & 85258 & $0.04$ & $0.06$     & A   \\
    $\gamma$ Sge   & 189319 & 98337 & $1.0$ & $10.0$     & C   \\
    $\psi$ Phe     & 11695  & 8837 & --- & ---     & C   \\
    $\epsilon$ Eri & 22049  & 16537 & $0.4$ & $0.9$     & A   \\
    Gmb 1830       & 103095 & 57939 & --- & ---     & C   \\
    61 Cyg A       & 201091 & 104214 & $1.0$ & $7.0$     & B   \\
    61 Cyg B       & 201092 & 104217 & $1.0$ & $7.0$     & B   \\ \hline
  \end{tabular}
\end{table}

\section{Results} \label{sec:results}
\subsection{Benchmark ages}
Based on all of the age information on each star (from the literature and this work), we have defined benchmark ages based on the discussions in Appendix~\ref{sec:appA}.
The ages are given in terms of a range, i.e. a lower and upper limit on the age, in \autoref{tbl:benchmark_ages}.
These ranges are our recommendations based on the combination of age estimates from the literature and this work, and they are generally conservative estimates in the sense that we have attempted to take into account the scatter in the ages related to both statistical and systematic uncertainties.
Because these ranges have been compiled from multiple sources that use different observations, techniques and models, we have not attempted to synthesise formal confidence intervals.
Our results should rather be interpreted as the ranges within which we are confident that the age of the star in question will lie, taking into account both systematic and statistical uncertainties.
In some cases we have not been able to set a well-defined upper limit (HD~140283, HD~22879, $\mu$~Cas, HD~220009, and $\alpha$~Tau), and in a few cases we give no limits at all (HD~122563, $\phi$~Phe, and Gmb~1830).
For the two binary systems with both components in the sample ($\alpha$~Cen and 61~Cyg), we have assigned the same age range.
This is mainly because they are expected to have been born together, but the independent age estimates in the literature do also agree well between the components for both systems.

In \autoref{tbl:benchmark_ages} we have also given each age range a rank of A, B, or C.
Rank A contains the stars for which the age range is well-defined and within an interval of a few Gyr.
Most of the stars with this rank are F or G~dwarfs and subgiants, but a few of them are young giants and one is the K~dwarf $\epsilon$~Eri which has a well-determined age based on gyro-/chromochronology.
Rank B is for stars with more uncertain ages or no upper age limit, but which still have some age information.
This includes stars with a large scatter between different age estimates and giants like $\mu$~Leo with broad $\mathcal{G}$~functions (see \autoref{fig:gfuncs_subset}c).
Finally, rank C is given to stars for which little to no age information has been obtained.
These are mostly giants for which model fitting gives little age information and the number of literature values is low.
For stars with rank A, the middle of the age range can be adopted if a single value for the age is desired.

\subsection{Special cases}
A few of these results deserve some comments here in addition to the discussion in Appendix~\ref{sec:appA}.
In the case of the 61~Cyg system, there is a tension between rotation/activity-based ages and model fitting ages.
The model fitting ages by \citet{2008A&A...488..667K} are based on the combined knowledge of the masses, radii, metallicities, temperatures, and luminosities which constrain the models well and give estimates of $6\pm1$~Gyr.
The rotation/activity-based ages by \citet{2007ApJ...669.1167B} and \citet{2008ApJ...687.1264M} are in the range 2--4~Gyr due to their different calibrations.
We give a conservative age range of 1--7~Gyr based on estimates from both methods; however, it is possible that gyrochronology is more reliable than model fitting in this case.
The reason is that \citet{2008A&A...488..667K} tested two different mass determinations for the components (see their section~4) and only with one of them could they obtain a satisfactory fit to the data.
At the same time, the ages based on gyrochronology are below the solar value where the activity-age relations are thought to be most reliable.
As an alternative to the range given in \autoref{tbl:benchmark_ages}, one can adopt 2--4~Gyr based on the results of gyrochronology.

In some cases (e.g. $\mu$~Cas) we use the fact that the rotation/activity-ages are above about 5--6~Gyr to exclude ages below about 3~Gyr.
The argument here is that if the star had an age of 3~Gyr or lower, and chromochronology is precise to within 60 per cent, then the derived age should not be above 5~Gyr.
In the case of Gmb~1830, however, we have not given any constraints on the age despite a number of age estimates from rotation/activity.
The majority of these age estimates are in the range 3--6~Gyr, but \cite{2004A&A...423..517R} give an age close to 13~Gyr based on a calibration of chromochronology which takes the metallicity into account.
For the low metallicity of this star ($-1.46$~dex) it is not clear how well the calibration of e.g. \citet{2008ApJ...687.1264M} performs since it is based on clusters with solar-like metallicities.
On the other hand, for $\mu$~Ara we find good agreement between age estimates based on model fitting and chromochronology from all sources except for \cite{2004A&A...423..517R}, which calls into question the reliability of their metallicity calibration of chromochronology.
Therefore, as an alternative to the very conservative lack of age information in \autoref{tbl:benchmark_ages} for Gmb~1830, one can adopt a range of 3--6~Gyr based on the rotation/activity-based estimates, but keep in mind that these relations are untested at the low metallicity of this star.

\section{Discussion} \label{sec:discussion}
\subsection{Reliability of the benchmark ages} \label{sec:discuss_reliability}
The quality of the age estimates given in \autoref{tbl:benchmark_ages} vary from well-defined intervals on the order of a gigayear to the complete lack of any constraints.
The majority of the stars with well-defined intervals, which have been assigned rank A, are subgiants and F and G~dwarfs.
Among these, a few stand out as having particularly reliable ages in the sense that many different estimates based on different methods are consistent with each other.
For $\alpha$~Cen~A the benchmark age is based on 25 more or less consistent literature values from model fitting, including some with asteroseismic constraints \citep{2012MNRAS.427.1847B, 2014A&A...566A..82L}, and gyrochronology with four independent calibrations \citep{2007ApJ...669.1167B, 2008ApJ...687.1264M, 2011MNRAS.413.2218D, 2015MNRAS.450.1787A}.
Similarly, 18~Sco and $\mu$~Ara both show great consistency between 25 or more literature ages based on model fitting (with and without asteroseismology) and gyrochronology.
Unfortunately, these stars represent only a small region in the HR~diagram where gyrochronology is at its most reliable since the star is on the main sequence and isochrone fitting works well since the star is hotter than the turn-off of the oldest isochrones.
Even with fewer literature values, the rest of the subgiants and F and G~dwarfs with rank A are likely also reliable due to their placement in the HR~diagram.

The ages of the giants are overall less reliable and only the youngest of them have been given rank A.
The problem is that the only method we have considered for these stars is model fitting without asteroseismology, and the number of literature estimates is lower than for the dwarfs (only about 1--5 per star).
Thus, only the stars located on the young isochrones, which are separated from the older ones, have precise age estimates.
An expansion of the benchmark sample to include more giants with asteroseismic data could increase the number of older benchmark giants with reliable ages.
Giant stars can be classified as either hydrogen shell or helium core burning based on the period spacings of their dipole oscillations \citep{2013ApJ...765L..41S, 2018MNRAS.476.3233H}.
Combining this information with precise surface parameters and asteroseismic observables will give more precise ages than what we achieved for the giants of the current sample.
Among the possible expansions of the sample discussed by \citet{2015A&A...582A..49H}, four \textit{Kepler} giants (HD~175955, 177151, 181827, and 189349) are mentioned.
We applied Bayesian isochrone fitting to these stars by adopting surface parameters from \citet{2012A&A...543A.160T} and \citet{2013MNRAS.434.1422M} and find that they are all relatively young (ages of below about 4~Gyr), but this is without including asteroseismic constraints.
They may still be valuable additions to the sample to increase the number of giants with precise benchmark ages.

What has been discussed here is to a large degree captured by the ranking we have assigned to each star.
Thus, we recommend that only the ages of the 16 stars with rank A are used for validation purposes.
For now this is a small sample which mainly includes young stars ($< 8$~Gyr) due to the difficulties of putting tight limits on the ages of older stars.
However, stars like HD~140283 and 22879, which we have assigned rank B due to the lack of an upper limit on the age, are certainly old and the lower limits could still be used for validation.

\subsection{Old metal-poor stars} \label{sec:old_mpoor}
A number of the most metal-poor stars in the sample fall off the old edge of our isochrone grids at their observed metallicity: HD~84937, HD~140283, HD~22879, $\mu$~Cas, HD~122563, and $\psi$~Phe.
This results in $\mathcal{G}$~functions which peak at the maximum age of each grid.

For some stars this problem is resolved when they are fitted to $\alpha$-enhanced isochrones according to the observed $\alpha$-abundances shown in \autoref{fig:alpha_abundances}.
This is the case for the subgiant HD~84937, for which the benchmark age is based on the results of model fitting to $\alpha$-enhanced isochrones, and for the dwarf HD~22879 which has an age between 13 and 14~Gyr in our $\alpha$-enhanced fit.
For the subgiant HD~140283 most of the literature ages are also at or above 14~Gyr; however, its position in the HR~diagram is so close to the 14~Gyr isochrone that it would take only a small change in the observed parameters, or in the isochrones, to move it below the age of the Universe.
For example, \citet{2015A&A...575A..26C} gives an age of 12~Gyr after assuming an extinction of $A_V = 0.1$~mag.
 The discrepancy is also minor for the dwarf $\mu$~Cas for which both \citet{2011A&A...530A.138C} and \citet{2014MNRAS.443..698S} give an age of about 6~Gyr.
 These ages are based on model fitting and therefore not necessarily reliable for this low-mass main sequence star; however, they do show that a difference in stellar parameters or models can place this star within the limits of the models.

The biggest discrepancies between the observations and models are seen for the two giants HD~122563 and $\psi$~Phe.
\citet{2015A&A...582A..49H} also noted this when deriving masses for the sample and they based their mass of $\psi$~Phe on models with a metallicity about 1~dex higher than the benchmark value.
It is not clear what causes these large discrepancies which is why we have chosen to give no ages for these two stars.

\begin{figure*}
  \center
  \includegraphics[width=0.9\textwidth]{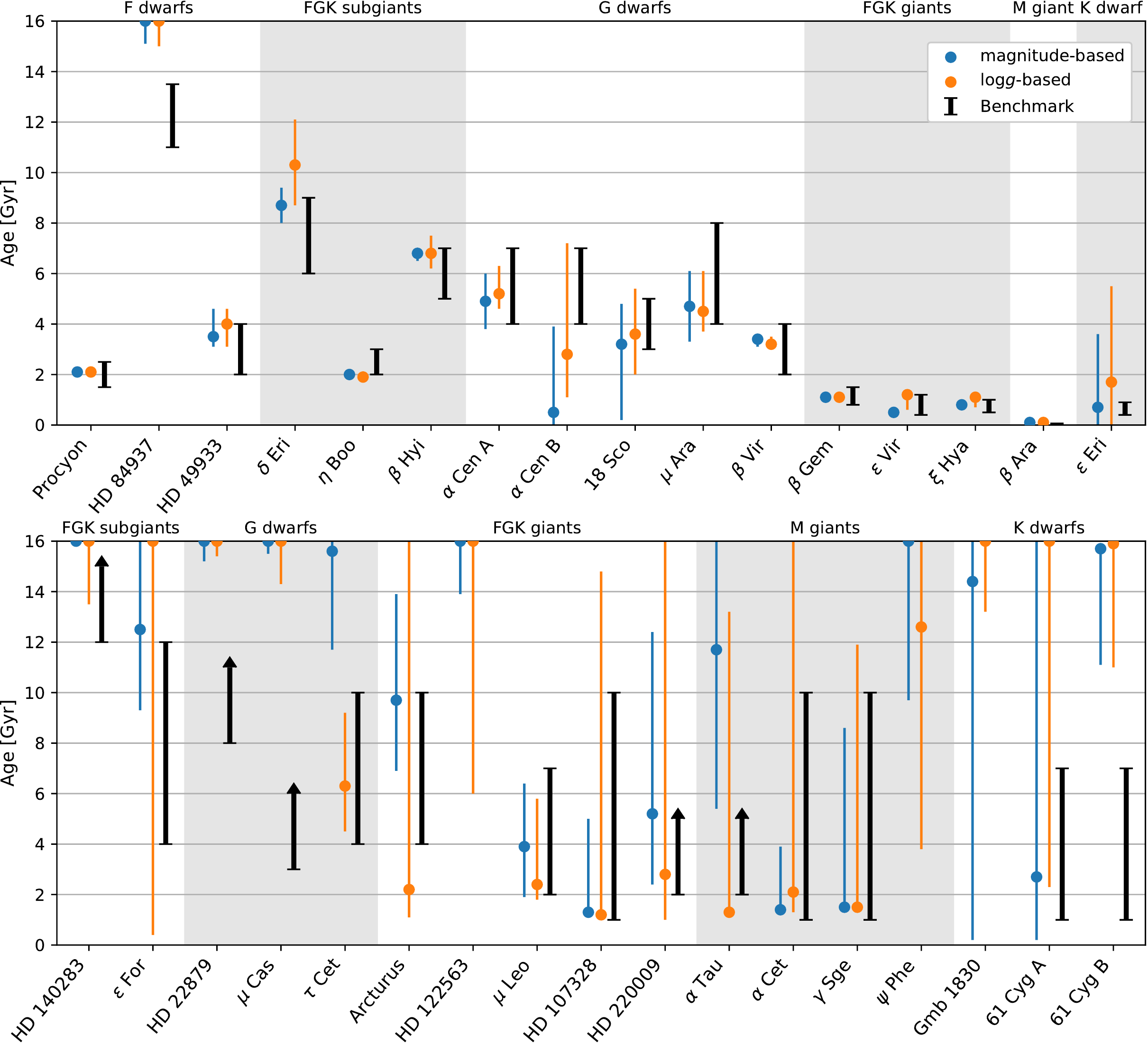}
  \caption{Benchmark ages compared with ages determined in this work based on Bayesian isochrone fitting to the MIST models.
    The ages and uncertainties for the isochrone-based results correspond to the mode and 1$\sigma$ interval of the $\mathcal{G}$~function as described in the final paragraph of section~\ref{sec:g_functions}.
  The upper panel shows all stars with rank A in \autoref{tbl:benchmark_ages} and the lower panel shows ranks B and C.
  The stars are sorted according to their classification as indicated on top of the panels and by the alternating shading of the background.}
  \label{fig:benchmark_comp}
\end{figure*}

\subsection{Comparison between benchmark and Bayesian ages}
Now we turn the attention towards the ages based on Bayesian isochrone fitting and discuss them with an outlook towards isochrone ages for large spectroscopic surveys.
For this purpose, the benchmark and Bayesian ages (from the MIST isochrones) are compared in \autoref{fig:benchmark_comp}.
The stars are divided by rank of the benchmark age and grouped by their spectral classification.
It is clear that the stars with rank A are also the ones with well-defined ages from isochrone fitting, with only a few exceptions.
This highlights a shortcoming of this comparison, namely that most of the benchmark ages are based in large part on age estimates from some sort of model fitting.
Therefore, the stars with well-defined benchmark ages are also predominantly the ones for which isochrone fitting works well (and vice versa).

However, there are a few cases where the benchmark values are not mainly based on model fitting.
For the K~dwarf $\epsilon$~Eri, the benchmark age is based on a large number of age estimates from gyro- and chromochronology.
The isochrone-based ages are much more uncertain, but the most likely ages are close to the benchmark value.
It should be mentioned that the good agreement between the benchmark and magnitude-based age for this star is slightly misleading since we only show the results based on the MIST isochrones.
The $Y^2$ and PARSEC isochrones give ages of 8 and 11~Gyr, respectively, simply due to slight differences in the location of the main sequence in the models.
The $\log g$-based ages, however, do not depend strongly on the choice of isochrones.
Thus, this is an example of a K~dwarf for which a very precise (and accurate) value of $\log g$ can be used to constrain the age, although not nearly as precisely as gyrochronology.
However, with an uncertainty of just $0.03$~dex in $\log g$, this is really the best-case scenario, and surface gravities from large spectroscopic surveys will not be precise enough to date a star like this (see the section below).
Another example is the $\alpha$~Cen system.
For $\alpha$~Cen~A the isochrone-based ages agree with the benchmark age, but for $\alpha$~Cen~B, which is further down the main sequence, the isochrone-based ages are both less precise and accurate.
Like for $\epsilon$~Eri, the $\log g$-based age is more reliable than the magnitude-based one owing to the very precise surface gravity.
Finally, $\tau$~Cet is another dwarf which can be dated with $\log g$ but not the magnitude.
For stars with higher surface gravities, i.e. the K~dwarfs in the lower panel of \autoref{fig:benchmark_comp}, even the precise benchmark parameters are not precise enough to give any age information from isochrone fitting.
This is seen clearly for the 61~Cyg system whose benchmark ages are based partly on gyrochronology.

For giants, only the youngest ones have reliable isochrone ages based on both $\log g$ and the magnitude, as discussed in Section~\ref{sec:discuss_reliability}.
For older giants, the ones with the most well-constrained benchmark ages are Arcturus and $\mu$~Leo for which only the magnitude-based isochrone ages are well-defined and consistent between the three different isochrone grids.
For giants with uncertain ages, there is a tendency for the $\mathcal{G}$~function to peak at an age of around 2~Gyr and fall off steadily towards the upper edge of the grid.
This is seen for HD~107328, $\alpha$~Cet, and $\gamma$~Sge (see the $\mathcal{G}$~functions in \autoref{fig:gfuncs_MIST} for reference).
What this means is that the single most probable age is low (around 2~Gyr); however, there is a significant probability that the star is in fact older.
If we had a large sample of these stars and adopted the most probable age as our estimate for all of them, we would underestimate the individual ages on average since the real ages are distributed like the $\mathcal{G}$~function of a single star.
This makes it difficult to assign an age for each individual star of this kind which is worth keeping in mind when deriving isochrone ages for large samples of giants.

To summarise this discussion, both the $\log g$-based and magnitude-based isochrone ages are well-defined for most of the F and G~dwarfs and subgiants, and the ones which are not well-defined are predominantly old.
Also the young giants can be dated by both $\log g$-based and magnitude-based fitting, but the older ones only by the magnitude, and even then it can be difficult to define the age based on the often very asymmetric $\mathcal{G}$~function.
Late G~dwarfs and early K~dwarfs can also be dated by fitting to $\log g$ when it is known very precisely, as is the case for the benchmark stars.
We expect that these general conclusions will carry over into isochrone-based age dating of large samples from spectroscopic surveys.
However, the stellar parameters will generally be less precise than what is available for the benchmark sample, this point is discussed in the following section.

\begin{figure}
  \center
  \includegraphics[width=\columnwidth]{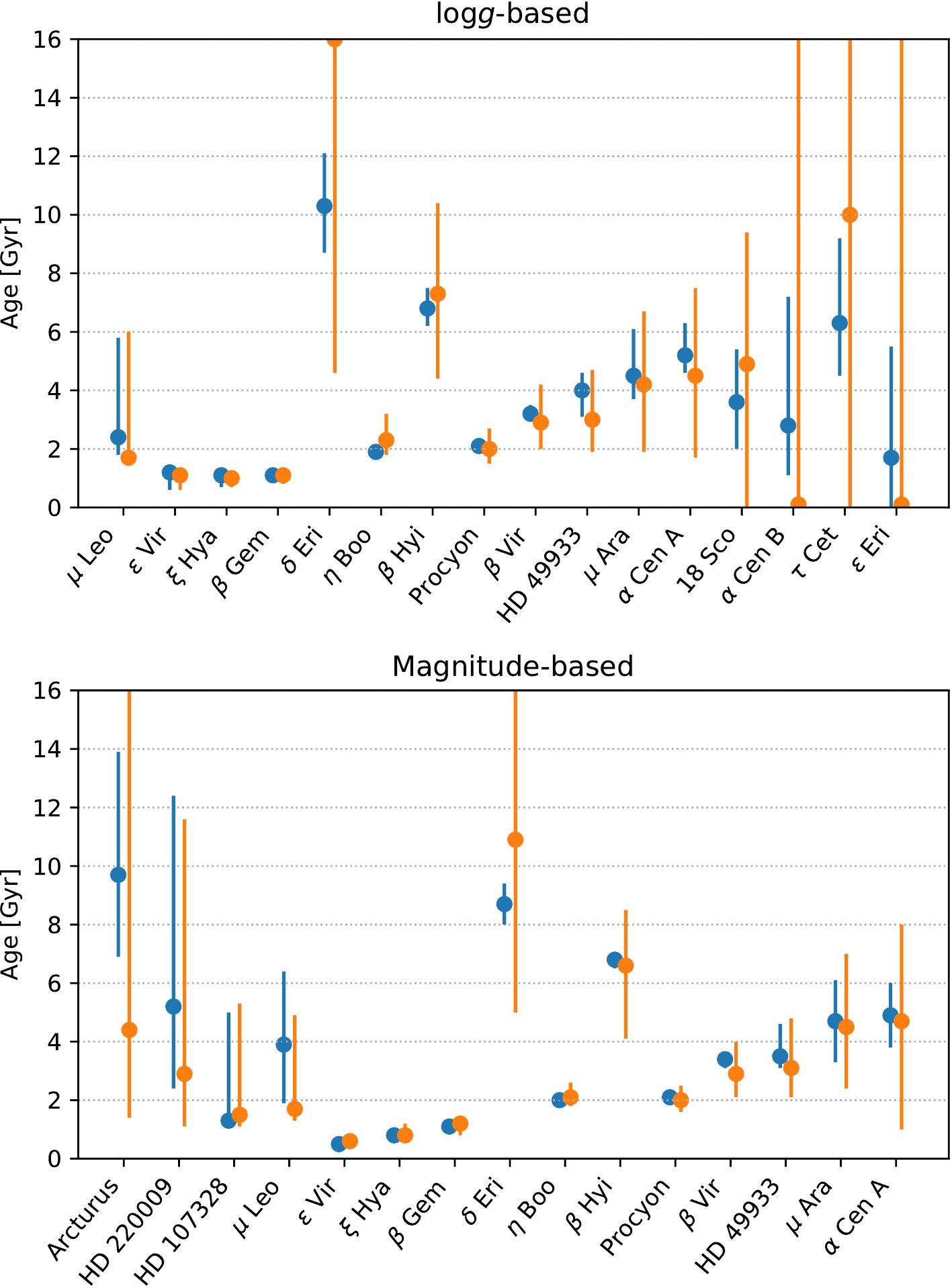}
  \caption{$\log g$-based ages (upper panel) and magnitude-based ages (lower panel) before and after
  increasing the uncertainties on the stellar parameters. For both sets of ages the uncertainties on
  $T_{\mathrm{eff}}$ and [Fe/H] have been increased to 150~K and 0.15~dex, respectively. The
  uncertainty on $\log g$ has been increased to 0.20~dex for the $\log g$-based results, and the
  uncertainty on the parallax has been increased to 20 per cent for the magnitude-based results. In both
  cases only the stars for which the age estimate is well-defined are shown which means that
  different stars are present in the two panels. The stars are sorted according to $\log g$, so the giants
  are on the left and the dwarfs are on the right.}
  \label{fig:MIST_unc}
\end{figure}

\subsection{Impact of stellar parameter uncertainties}
With the precise benchmark parameters adopted in this study (\autoref{tbl:obs_data}), the ages derived from isochrone fitting are precise to better than 5 per cent in the best cases.
This means that the benchmark ages given in \autoref{tbl:benchmark_ages} for stars like Procyon, $\eta$~Boo, and $\beta$~Vir are limited by the systematic scatter seen in the literature due to the use of different models and values of the stellar parameters.
Stellar parameters from large spectroscopic surveys will in general not be as precise as the benchmark values.
Additionally, the benchmark stars have very precise parallaxes (most with relative uncertainties below 5 per cent) since they are all relatively nearby, and this will not be the case for more distant stars in the surveys.

For the stars with well-defined isochrone ages, we have tested the impact of increasing the uncertainties on all of the stellar parameters to be more in line with those coming from large spectroscopic surveys.
We have taken a slightly pessimistic approach and increased the uncertainties on the stellar parameters to 0.15~dex in [Fe/H], 150~K in $T_{\mathrm{eff}}$, 0.2~dex in $\log g$, and 20 per cent in parallax\footnote{For reference, in the second data release of GALAH, the typical uncertainty is at or below 0.08~dex in [Fe/H], 100~K in $T_{\mathrm{eff}}$, and 0.20~dex in $\log g$.}.
Stars which already had uncertainties at or above these values have not been changed.

\autoref{fig:MIST_unc} shows the ages before and after uncertainty inflation in order of increasing $\log g$ from left to right.
Starting with the dwarfs and subgiants, the impact of increasing the uncertainties in the stellar parameters increases when going towards higher surface gravities and ages.
This is to be expected since the isochrones lie closer in the HR~diagram for both higher surface gravity and age.
The most precise ages have a relative uncertainty of about 25 per cent compared to 5 per cent before increasing the uncertainties.
The increase in uncertainties has a large effect for the old subgiant $\delta$~Eri where the upper limit on the age is lost.
For the dwarfs with the highest surface gravities, which could only be dated with $\log g$ before increasing the uncertainties, all age information is lost after the increase.
18~Sco, with a surface gravity of 4.44~dex, seems to mark the transition between well-defined and ill-defined ages.
So for $\log g \gtrsim 4.4$~dex isochrone fitting with survey-like parameter uncertainties is no longer useful for age determination; however, this limit will shift slightly depending on the metallicity which is solar in the case of 18~Sco.

For giants, the young ones ($\epsilon$~Vir, $\xi$~Hya, and $\beta$~Gem) still have precise ages after inflation.
The ages of the older ones, which can only be dated using the magnitude, become more uncertain with the most uncertain one being the oldest (Arcturus).
At the same time, the $\mathcal{G}$~functions tend towards a shape with the mode at a low age as discussed in the previous section.
Thus, it will be difficult to reliably date old giants in large surveys without additional constraints from e.g. asteroseismology.

\section{Conclusions} \label{sec:conclusions}
We have investigated the ages of the 33 \textit{Gaia} benchmark stars both by reviewing the literature and deriving new sets of ages based on Bayesian isochrone fitting.
The literature ages are mainly based on model fitting, which for some stars includes asteroseismic data, but we have also found estimates based on rotation or chromospheric activity for many of the dwarfs.
Based on the ages we have found and derived, we have defined benchmark ages for the stars in the form of lower and upper limits which take into account both statistical and systematic uncertainties in the different age dating methods.

For 16 out of the 33 stars we believe our benchmark ages are reliable and suitable for use as validation of other age determinations.
When using these ages, we recommend taking the middle of the interval given in \autoref{tbl:benchmark_ages} as the most likely age, and interpret the range as the combination of statistical and systematic uncertainties related to the use of different models and methods.
Even for the subgiants and turn-off stars with the most precise isochrone-based age estimates -- such as $\beta$~Hyi, Procyon, and $\eta$~Boo -- the benchmark age cannot be constrained to better than within 1~Gyr due to systematic uncertainties related to the use of different stellar models.

The ages of the remaining 17 stars are either less precise, only have a lower limit, or are not known at all.
Among these, the metal-poor giants HD~122563 and $\phi$~Phe fall completely outside our grids of isochrones.
Other metal-poor stars fall only slightly outside of our grids at the old edge.
Most of these stars are $\alpha$-enhanced and we find that they are brought within the edges of the grids when this is taken into account in the models.
We also tested fitting to the surface metallicity of each model, which is affected by diffusion, instead of using the initial value of each isochrone.
The effect of this is minor except for the turn-off star HD~49933 which becomes more than 1~Gyr younger.

By comparing our isochrone-based age estimates with the benchmark values, we find that isochrone fitting can reliably date the 16 stars with well-defined ages; however, this is partly due to the inclusion of model fitting ages in the definition of the benchmark ages.
With the very precise surface gravities of the benchmark dwarfs, the $\log g$-based ages can even be used to date late G~dwarfs and early K~dwarfs; for example $\alpha$~Cen~B and $\epsilon$~Eri.
But after increasing the uncertainties on the stellar parameters to be closer to those from large spectroscopic surveys, we find that isochrone fitting gives no age information for stars with $\log g \gtrsim 4.4$~dex.
Using the benchmark results as an outlook towards isochrone-based ages in large spectroscopic surveys, we should be able to determine reliable ages for all dwarfs and subgiants with $\log g \lesssim 4.4$ (although the oldest ones will only have lower limits) as well as for the youngest (most massive) giants.
Ages derived for older giants should be analysed with care due to the often asymmetric and very extended $\mathcal{G}$~functions.

The 16 stars with reliable ages are mostly subgiants and F and G~dwarfs.
They also include four young giants (ages $<2$~Gyr) and the young K~dwarf $\epsilon$~Eri which has been dated based on gyro- and chromochronology.
A future expansion of the sample, especially including giants with asteroseismic observations, would be helpful in increasing the number of benchmark stars in different evolutionary stages with well-known ages.

\section*{Acknowledgements}
C.L.S, S.F., and R.C. were supported by the project grant `The New Milky Way' from the Knut and Alice Wallenberg foundation.
C.L.S. and S.F. were  supported by the grant 2016-03412 from the Swedish Research Council.
L.L. was supported by the Swedish National Space Board.
R.C. was supported by funds from the eSSENCE Strategic Research Environment.
This research has made use of the SIMBAD database, operated at CDS, Strasbourg, France




\nocite{1998A&A...329..943N}\nocite{1999A&A...348..897L}\nocite{2001A&A...371..943C}\nocite{2003A&A...409..251M}\nocite{2004MNRAS.349..757L}\nocite{2004A&A...426..309G}\nocite{2004A&A...425..187T}\nocite{2004A&A...413..251K}\nocite{2007A&A...468..663T}\nocite{2013ApJ...769....7L}\nocite{2014ApJ...787..164G}\nocite{2006A&A...451.1065G}\nocite{2005A&A...436..253T}\nocite{2004A&A...413.1045G}\nocite{1999A&A...348..487R}\nocite{2001A&A...367..253F}\nocite{2005ApJ...635..547G}\nocite{2005A&A...434.1085C}\nocite{1998A&A...330.1077D}\nocite{2010A&A...519A..87B}\nocite{2008A&A...488..653P}\nocite{2012A&A...542A..84D}\nocite{2012A&A...538A..21S}\nocite{2016ApJ...833..161G}\nocite{2010A&A...521A..12M}\nocite{2014ApJ...785...33S}\nocite{2004A&A...426..601D}\nocite{2015ApJ...803...90P}\nocite{2011ApJ...734...70A}\nocite{2009A&A...501..687D}\nocite{2010A&A...519A.101D}\nocite{2010A&A...512L...5S}\nocite{2012ApJ...746..143L}\nocite{2012A&A...546A..83L}\nocite{2014A&A...572A..48R}\nocite{2014A&A...563A..52P}\nocite{2016A&A...593A..65N}\nocite{2016A&A...593A.125S}\nocite{2015A&A...579A..52N}\nocite{2016A&A...591A..89M}\nocite{2016A&A...590A..32T}\nocite{2003AJ....125.2664L}\nocite{2015A&A...575A..18B}\nocite{2016A&A...585A...5B}\nocite{2005A&A...443..609S}\nocite{2013ApJ...768...25G}\nocite{2010A&A...513A..49S}\nocite{2016A&A...585A..73N}\nocite{2014ApJ...794..159B}\nocite{2000ApJ...533L..41S}\nocite{2001ApJ...561.1095B}\nocite{2017MNRAS.465.2734M}

\bibliographystyle{mnras}
\bibliography{references}



\appendix
\section{Discussions of individual stars} \label{sec:appA}
\begin{figure*}
  \center
  \includegraphics[width=\textwidth]{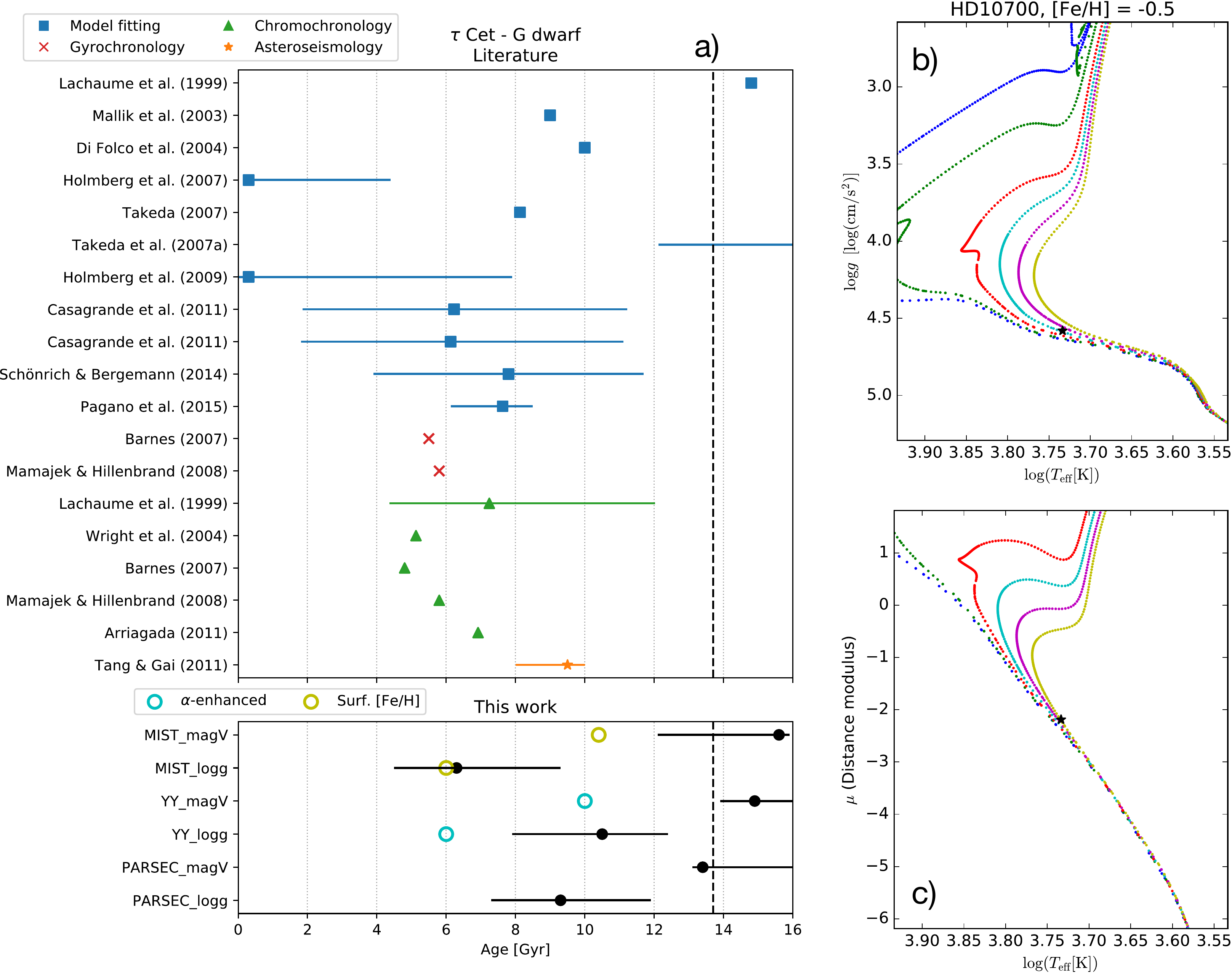}
  \caption{Ages and HR~diagrams for the star $\tau$~Cet (HD~10700).
  a) Ages collected from the literature (top panel), and ages determined in this work (bottom panel).
  The different methods used in the literature are indicated with different colours and symbols.
  The coloured open circles in the bottom panel show the special cases of fitting to $\alpha$-enhanced $Y^2$ isochrones (when relevant) and fitting to the current surface metallicity of the MIST isochrones.
  Uncertainties on the ages are plotted for all of the literature values for which they were available, and for all ages determined in this work; however, they may be smaller than the symbol size in both cases.
  The vertical dashed line is to indicate the age of the Universe of 13.7~Gyr as determined by WMAP \citep{2013ApJS..208...20B}.
  b) Location of the star in ($T_\mathrm{eff}$, $\log g$)-space (star symbol) with MIST isochrones of the given metallicity and ages of 0.5, 1, 3, 6, 10, and 15~Gyr. 
  c) The same as b), but in ($T_\mathrm{eff}$, distance modulus)-space where the observed distance modulus is based on the parallax, and the distance modulus of the models is based on the observed $V$ magnitude.}
  \label{fig:ages_HD10700}
\end{figure*}

Here we discuss, for each star individually, the ages found in the literature and from our own fits to the different isochrones and observables ($V$ band magnitude and $\log g$).
For the stars for which the adopted values of $T_\mathrm{eff}$ and $\log g$ have not been taken from Table~10 in \citet{2015A&A...582A..49H}, their sources are listed in the discussion.
For each star, the ages found in the literature and derived in this work are summarised, and comments are made on the overall agreement of the estimates and on outliers and other peculiarities.
We also give our conclusion on the age of the stars based on the data we have available here.
Each discussion is accompanied by a figure which shows all of the ages found in the literature and this work, as well as the position of the observed stellar parameters relative to the MIST isochrones in both ($T_\mathrm{eff}$, distance modulus)-space and ($T_\mathrm{eff}$, $\log g$)-space.
The complete set of these figures can be found in the online material and an example for $\tau$~Cet (HD~10700) is given in \autoref{fig:ages_HD10700}.

\subsection*{Procyon}
\paragraph*{Literature}
All the literature age values based on model fitting are very precise since Procyon is at the main sequence turn-off where the isochrones are well separated, and the ages fall around 2~Gyr with a low amount of scatter.
It seems that \citet{2015ApJ...804..146D} did not include a metallicity constraint on their fit which would explain the relatively large uncertainty on their estimate.
The five age estimates based on model fitting with asteroseismic constraints agree with the rest of the
literature, but they do not pin down the age more accurately.
The solar-like oscillations of Procyon have been difficult to characterise and the two estimates given by \citet{2010AN....331..949D} are based on two different scenarios for the identification of the oscillation modes (see \citet{2010ApJ...713..935B} for the details).
The one estimate based on chromochronology agrees on an age around 2~Gyr, while the one based on the X-ray luminosity gives an age of around 4~Gyr.
The deviation of the X-ray age from the rest of the literature is not very concerning since the X-ray ages are the least precise method we consider.
\paragraph*{This work}
The age estimates from this work agree with the literature and they are all slightly above 2~Gyr.
We find no difference between the ages determined with different isochrones which helps explain the very low scatter of the literature values.
\paragraph*{Conclusion}
Both the literature and our own estimates agree on an age close to 2~Gyr with little scatter.
Based on the spread of the literature values we give the age as 1.5--2.5~Gyr.

\subsection*{HD 84937}
\paragraph*{Literature}
All of the literature values are based on model fitting, and most of them put the age of this star above 12~Gyr.
The two lowest values are both based on fitting to $\log g$ instead of the magnitude.
The $\log g$ value used by \citet{2015ApJ...804..146D} is particularly high ($4.46\pm0.14$~dex compared to the benchmark value of $4.06\pm0.04$~dex), and they make no mention of including a constraint on the metallicity in their fit.
These factors are likely enough to explain the low age they find.
\citet{2014ApJ...792..110V}, who find an age of 12~Gyr, discuss the fact that at low metallicities, the location of the turnoff is more sensitive to the oxygen abundance than to the abundance of iron.
They determined a value of [O/Fe] $= 0.44$~dex and included oxygen enhancement in their models; by neglecting this in the models one may overestimate the age (however, see the discussion of HD~140283 where oxygen enhancement does not seem to bring the age below that of the Universe).
\paragraph*{This work}
Most of the age estimates from this work are at the upper edge of the isochrone grids since this star seems to be cooler than the oldest isochrones.
This is the case for the MIST and PARSEC ischrones; however, the star is slightly hotter than the oldest $Y^2$ isochrone which gives us a single well-defined age estimate of about 13.5~Gyr when fitting to the magnitude.
This star has one of the highest levels of $\alpha$-enhancement within the sample, and when fitting to $Y^2$
isochrones with [$\alpha$/Fe] = 0.4 both the magnitude and $\log g$-based ages become well-defined
at around 12~Gyr.
This is in agreement with the value found by \citet{2014ApJ...792..110V} using oxygen-enhanced
models.
We also see a slight shift towards lower ages when fitting to the current surface metallicity of the MIST
isochrones.
\paragraph*{Conclusion}
The most reliable literature values are all at 10~Gyr or above, and all our own estimates are above
13~Gyr.
However, when including $\alpha$-enhancement in the models, the age is lower (and well-defined with
upper confidence limits below the age of the Universe) as found in our own results and by
\citet{2014ApJ...792..110V}.
We take the estimates based on $\alpha$-enhanced models to be the most reliable and give the age as
11.0--13.5~Gyr.

\subsection*{HD 49933}
\paragraph*{Literature}
All ages are based on model fitting and they are very precise since this star is located right at the turn-off where isochrone ages are most reliable.
Despite the precise ages, they span a relatively large range of 2--4~Gyr, and the estimates that included asteroseismic data show a slight preference for the higher end of this interval with one of them at $4.4$~Gyr \citep{2009A&A...506..235B}.
The two lower asteroseismic ages \citep{2011A&A...534L...3B, 2011A&A...532A..82O} are based on more detailed frequency information since they include the small frequency separations in addition to the large ones. 
\citet{2011A&A...534L...3B} additionally include the stellar radius derived from an interferometric determination of the angular diameter as a constraint.
The four lowest values are all from different versions of the GCS which means they are correlated (in fact two of them are equal \citep{2007A&A...475..519H, 2009A&A...501..941H}).
\paragraph*{This work}
The ages from this work show good agreement with the literature values and mainly fall within the interval 3--4~Gyr.
There is no significant difference between fitting to $\log g$ or the magnitude, but there is a slight difference of up to about 1~Gyr between the ages determined using different isochrones.
So in this case (a turn-off star with precise stellar parameters) the models limit the accuracy of our age estimate.
Additionally, we see a significant shift in the age when fitting to the current surface metallicity instead of the initial one, suggesting that the age is overestimated when this effect is neglected.
\paragraph*{Conclusion}
The literature suggests an age of 2--4~Gyr, and our values vary within the same interval depending on the adopted isochrones and whether we fit to the current or initial surface metallicity.
The scatter of our own values suggests that the scatter in the literature may be caused by the choice of isochrones and perhaps the way the observed metallicity is compared with the models.
Even though individual age estimates are very precise, we are limited by systematics leading us to give the age as 2--4~Gyr.

\subsection*{$\delta$ Eri}
\paragraph*{Literature}
The ages based on model fitting agree well with each other, and most values are around 6~Gyr.
The age given by \citet{2014A&A...562A..71B} is the only one based on $\log g$ instead of the magnitude, and the uncertainty on $\log g$ explain the larger uncertainty on this age.
The single age estimate based on chromospheric activity \citep{2004A&A...423..517R} agrees with the isochrone ages even though it is less reliable since this star has evolved off the main sequence.
\paragraph*{This work}
The age estimates based on $\log g$ are higher than those based on the magnitude.
This star is an old subgiant, meaning that the isochrones are not as well separated as they are for young subgiants, so an age difference is easily introduced when using different observables.
Since the isochrones are better separated in magnitude than $\log g$, we believe the magnitude-based results are more reliable.
When fitting to the magnitude, the age we find with the $Y^2$ isochrones is in good agreement with the literature values.
However, with PARSEC and MIST isochrones we find ages about 2~Gyr higher which is difficult to explain since the literature values are based on a number of different models.
For example, \citet{2016A&A...588A..98M} also use PARSEC isochrones, but their age is about 3~Gyr lower than our PARSEC age.
Part of this difference may be explained by the benchmark temperature which is about 150~K lower than the typical temperature used in the literature.
For a discussion of this temperature difference, see \citet[section 5.2.3]{2015A&A...582A..49H}.
\paragraph*{Conclusion}
Most of the literature values agree on an age of 6~Gyr and are based on a number of different models.
When using the magnitude, the age estimates of this work span 7--9~Gyr; this is higher than the literature even when we use the same models.
The difference is possibly due to the higher spectroscopic temperatures used in the literature.
Overall, when considering the magnitude-based ages which we deem more reliable than the $\log g$-based, we arrive at an age of 6--9~Gyr depending on the exact temperature of the star and the adopted models.

\subsection*{HD 140283}
\paragraph*{Notes on input parameters}
The effective temperature determined by \citet{2015A&A...582A..49H} was not recommended for use as a reference value.
Instead, we adopted the mean spectroscopic literature value \citep[Table 11]{2015A&A...582A..49H}.
\paragraph*{Literature}
This star is a very metal-poor and old subgiant.
The literature values consistently place it at or above the age of the Universe with good precision owing to its location on the subgiant branch.
The lowest age is based on a higher metallicity of [Fe/H]~=~$-1.96$~dex \citep{2002A&A...394..927I}.
The age given by \citet{2014A&A...562A..71B} is the only one using $\log g$ instead of the magnitude, and the uncertainty on $\log g$ explain the larger uncertainty on this age.
The two values by \citet{2015A&A...575A..26C} differ in the adopted extinction value; they used $A_V=0.1$ to get the lower age, and $A_V = 0$ for the higher age.
\citet{2014ApJ...792..110V} discuss the fact that at low metallicities, the location of the turnoff is more sensitive to the oxygen abundance than to the abundance of iron.
They determined a value of [O/Fe] $= 0.64$~dex; by neglecting this in the models one may overestimate the age.
However, both \citet{2014ApJ...792..110V} and \citet{2013ApJ...765L..12B} apply oxygen-enriched models and still find that the age of the star is slightly higher than the age of the Universe.
\paragraph*{This work}
All of our age determinations hit the upper edge of the grids since the observed temperature is lower than the oldest isochrones at the observed metallicity.
Even though this star is so metal-poor, it is not significantly enhanced in any of the $\alpha$-elements measured by \citet{2015A&A...582A..81J} (see \autoref{fig:alpha_abundances}), so we have not fitted it to $\alpha$-enhanced models.
\paragraph*{Conclusion}
All of the ages from both the literature and this work are at or close to the age of the Universe.
In the literature, the lowest ages have been found based on a higher metallicity \citep{2002A&A...394..927I} or by adopting an extinction of $A_V=0.1$ \citep{2015A&A...575A..26C}.
In any case, the age must be close to that of the Universe and we give it as $>12$~Gyr.

\subsection*{$\epsilon$ For}
\paragraph*{Notes on input parameters}
The $\log g$ value determined by \citet{2015A&A...582A..49H} was not recommended for use as a reference value.
Instead, we adopted the mean spectroscopic literature value \citep[Table 11]{2015A&A...582A..49H}.
\paragraph*{Literature}
Most of the literature values are based on model fitting, and they are scattered approximately in the range 4--12~Gyr.
Interestingly, this large scatter is inconsistent with the relatively low uncertainties reported for many of these values and we have not been able to determine the source of this scatter.
It is too large to be caused by isochrone differences, and taking the values of \citet{2006A&A...458..609D} and \citet{2011A&A...530A.138C} as examples, their adopted stellar parameters are consistent within the uncertainties and also consistent with the benchmark values.
The one age based on chromochronology is around 9~Gyr, but the activity-age relation has not been calibrated for subgiants such as this star.
\paragraph*{This work}
Based on the two HR~diagrams, we see that the adopted $\log g$ is too high, and our fits with $\log g$ give no age information.
We adopted a value of $\log g = 4.07\pm0.30$~dex from the mean of spectroscopic literature values, but the benchmark value of $3.52\pm0.08$~dex is lower and more consistent with the isochrones.
Our fits using the magnitude give ages in the range 10--13~Gyr, but with uncertainties which reach the upper edges of the grids.
The isochrones are slightly different at the turn-off resulting in the lower age found using the $Y^2$ grid.
This is one of the most $\alpha$-enhanced stars in the sample, and when fitting to $\alpha$-enhanced $Y^2$ isochrones we find a significantly lower age (about 6~Gyr) which also has well-defined uncertainties.
The large impact of $\alpha$-enhancement for this star may explain some of the scatter in the literature, and it indicates that our own ages are overestimated with the MIST and PARSEC isochrones.
\paragraph*{Conclusion}
The ages from this work based on the magnitude indicate that this star is older than 10~Gyr, but when we include $\alpha$-enhancement in the $Y^2$ isochrones we get a value close to 6~Gyr.
The literature values are also very scattered, and since we cannot say for sure whether this is due to the treatment of $\alpha$-enhancement, we give the age conservatively as 4--12~Gyr.

\subsection*{$\eta$ Boo}
\paragraph*{Literature}
All ages are based on model fitting, and they are all very precise since this star is located right at the turn-off where isochrone ages are most reliable.
The majority of the ages fall in the interval 2--3~Gyr, and the estimates that include asteroseismic data prefer a value of around 2.5~Gyr with very high precision (only the age given by \citet{2004ApJ...612..454G} did not come with an uncertainty, the rest have uncertainties smaller than the symbol size).
The scatter of the values does not show an obvious correlation with the parameters ($T_{\mathrm{eff}}$ and [Fe/H]) which have been used in the fit.
Instead, the scatter may be caused by the star's placement right on the hook of the 3~Gyr isochrone.
For the fitting algorithms which weight each model according to the speed of evolution, the 3~Gyr hook solution is given a lower weight than a solution at a slightly lower age which is still making its way off the main sequence and towards the hook.
This explanation is consistent with the fact that the literature ages seem to shift towards lower values between the years of 2005 and 2007 when the Bayesian fitting method, which applies such model weights, became widely used.
\paragraph*{This work}
The ages from this work are all around 2~Gyr regardless of the adopted isochrones and whether we fit to the magnitude or $\log g$.
Like for some of the literature values, the uncertainties on our ages are smaller than the size of the symbols.
We tested the claim that the scatter of the literature values is due to the use of different fitting algorithms by turning off the model weights in our fit.
This resulted in an age of 3~Gyr in agreement with the older literature values, indicating that the different algorithms are indeed the cause of most of the scatter in the ages of this star.
\paragraph*{Conclusion}
The literature suggests an age of 2--3~Gyr, and the values from this work are all very close to 2~Gyr.
The scatter in the literature seems to be caused by the ambiguity between a 3~Gyr old star right at the hook and a 2~Gyr old star at the turn-off (but below the hook).
Bayesian fitting algorithms favour the younger solution since it is in a slower phase of evolution.
Considering that we cannot with certainty exclude the 3~Gyr solution, and the fact that the asterosesmic ages are around 2.5~Gyr, we give the age as 2--3~Gyr.

\subsection*{$\beta$ Hyi}
\paragraph*{Literature}
All ages are based on model fitting and they show good agreement since this star is located right at the turn-off where isochrone ages are most reliable.
The ages span values of 5--7~Gyr, and the estimates that include asteroseismic data prefer the higher end of the interval.
One outlier is present at an age of 3.5~Gyr \citep{2011A&A...532A..20F}, and it is not clear what causes the low age since it is based on stellar parameters consistent with those adopted in the rest of the literature.
The two asteroseismic ages determined by \citet{2011A&A...527A..37B} differ by the omission of microscopic diffusion in the models used to determine the lower age.
\paragraph*{This work}
The ages from this work show good agreement with the literature values.
There is no significant difference between fitting to $\log g$ or the magnitude, but there is a statistically significant difference of about 1~Gyr between the ages determined using different isochrones, with the $Y^2$ results being lowest.
So in this case, a turn-off star with precise stellar parameters, the models limit the accuracy of our age estimate.
\paragraph*{Conclusion}
The literature suggests an age of 5--7~Gyr, and our estimates vary between 6--7~Gyr depending on the adopted isochrones.
Thus, we give the age as 5--7~Gyr to account for the scatter in the literature.

\subsection*{$\alpha$ Cen A}
\paragraph*{Literature}
We have found an almost equal number of ages based on model fitting and rotation/activity-age relations.
Most of the model fitting ages are consistent with the range 4--5~Gyr, but two of them are higher at 8~Gyr.
The cause of this difference is not obvious since there is no significant difference in the adopted stellar parameters.
In fact, \citet{2007ApJS..168..297T} take their stellar parameters from \citet{2005ApJS..159..141V} which narrows down the source of the age difference to the choice of stellar models or algorithms applied in the fit.
We have also found two fits with asteroseismic constraints; the more precise of the two included both small and large frequency separations (see \citet{2012MNRAS.427.1847B} for the details) and gives an age of $4.8\pm0.5$~Gyr.
More results with asteroseismology exist in the literature based on simultaneous modelling of $\alpha$ Cen A and B to give an age of the system.
Such estimates generally put the age of the system in the range 5--7~Gyr \citep{2002A&A...392L...9T, 2003A&A...402..293T, 2004A&A...417..235E, 2004ApJ...600..419G, 2005A&A...441..615M}.
The ages based on gyrochronology are consistent with each other and with the ages based on model fitting, and they are based on five different calibrations of the method.
The highest age, found by \citet{2011MNRAS.413.2218D}, is based on the assumption that the period decays with time as $P\propto t^{0.5}$.
They find that this assumption overestimates ages compared to asteroseismology, and their second age estimate is based on the relation $P\propto t^{0.56}$ which they find by calibrating to a set of asteroseismic ages.
Finally, we found two estimates based on two different calibrations of chromochronology, and they agree with the rest of the literature.
Overall, the literature values agree well on an age in the range 4--7~Gyr.
\paragraph*{This work}
The age estimates of this work fall in the range 4--7~Gyr when including the uncertainties.
There is no significant difference between the ages based on the magnitude or $\log g$, but the choice of isochrones introduces a difference of almost 2~Gyr.
This is seemingly due to differences in the isochrones around the turn-off; for the 6~Gyr isochrones at this metallicity, the PARSEC turn-off is completely smooth whereas the both the YY and MIST turn-off shows the hook-like feature indicative of a convective core.
\paragraph*{Conclusion}
Based on a large number of age estimates in the literature from both model fitting (with and without asteroseismology) and rotation/activity-age relations, as well as our own estimates, we give the age of $\alpha$~Cen~A as 4--7~Gyr.

\subsection*{HD 22879}
\paragraph*{Literature}
Most of the literature values are based on model fitting, but they are very scattered and many have large uncertainties due to this star's location near the main sequence.
Many of the estimates are either above 12~Gyr, or have uncertainties which extend above 12~Gyr.
The value by \citet{2014A&A...562A..71B} stands out among the most recent age determinations with a value around 6~Gyr.
This is the only estimate based on $\log g$ instead of the magnitude.
The three estimates based on chromochronology are based on two different calibrations and give ages in the range 4--6~Gyr.
It is difficult to say whether these ages are more reliable than those based on model fitting since the activity-age relations are poorly calibrated beyond the solar age and for non-solar metallicities.
Still, the fact that they indicate a relatively high age most likely excludes a low age of $\lesssim 3$~Gyr.
\paragraph*{This work}
All age estimates from this work are at the upper edges of the isochrone grids.
The reason is clear based on the location in the HR~diagram; the star is cooler than the oldest isochrones, both when using the magnitude and $\log g$.
This is one of the stars with the highest $\alpha$-enhancement of the sample, and when fitting to $Y^2$ isochrones with [$\alpha$/Fe] = 0.4 both the magnitude and $\log g$-based ages decrease to a value close to 13~Gyr.
\citet{2014A&A...562A..71B} found a temperature and $\log g$ about 100~K and 0.25~dex higher than the benchmark  values, both of which pulls the age towards a lower value than our $\log g$-based ages.
They also included $\alpha$-enhancement in the models.
\paragraph*{Conclusion}
Most of the age estimates in the literature based on model fitting indicate that this is an old star, and most of our age estimates are at the upper edges of the grids.
The inclusion of $\alpha$-enhancement in the $Y^2$ isochrones lowers our age estimate below the upper grid edge, but it is still 13~Gyr.
We choose to weigh the position of this star in the HD-diagram higher than the chromochronology ages and give the age as $>8$~Gyr to allow some room for systematics in the isochrones or observables.

\subsection*{$\mu$ Cas}
\paragraph*{Notes on input parameters}
The $\log g$ value determined by \citet{2015A&A...582A..49H} was not recommended for use as a reference value.
Instead, we took $\log g$ from an estimate using the dynamical mass of the star \citep[section~5.4.1]{2015A&A...582A..49H}.
\paragraph*{Literature}
The ages based on isochrone fitting span the entire possible range and have high uncertainties.
This is due to this star's position on the main sequence where all of the isochrones overlap.
Two of the ages are very high and these are based on fits with temperatures consistent with the benchmark value \citep{2007PASJ...59..335T, 2015A&A...580A.111L}.
For the rest of the literature ages, which are all lower, either the isochrones have been shifted in temperature to align the main sequence with observations \citep{2007A&A...475..519H, 2009A&A...501..941H}, or an effective temperature about 150~K higher has been used to determine the age \citep{2011A&A...530A.138C}.
A few estimates based on rotation and chromospheric activity are available suggesting an intermediate age of around 5--6~Gyr; however, these relations are poorly calibrated beyond the solar age and for non-solar metallicities.
Still, the fact that they indicate a relatively high age most likely excludes a low age of $\lesssim 3$~Gyr.
\paragraph*{This work}
All of the age estimates from this work hit the upper edges of the grids because the star is cooler than the main sequence of the isochrones at the observed metallicity.
We can give no age estimate or limits based on these results.
\paragraph*{Conclusion}
Since model fitting is unreliable for this star, we base our age estimate on the rotation/activity values and give it as ${>3}$~Gyr.

\subsection*{$\tau$ Cet}
\paragraph*{Notes on input parameters}
The $\log g$ value determined by \citet{2015A&A...582A..49H} was not recommended for use as a reference value.
Instead, we took $\log g$ from an estimate based on asteroseismology \citep[Table 12]{2015A&A...582A..49H}.
\paragraph*{Literature}
The ages based on isochrone fitting span the entire possible range and have high uncertainties.
This is due to this star's position on the main sequence where all of the isochrones overlap.
One of the fits is based on asteroseismic data and gives an age of 8--10~Gyr \citep{2011A&A...526A..35T}.
However, a closer inspection of their results (their Figure 2) show that their observed oscillation frequencies match almost as well to a model with age 4~Gyr as to a model with age 8~Gyr ($\chi^2$ of the fit to the model frequencies is about 1.22 and 1.21, respectively, for the two ages).
The seven age estimates based on rotation/activity are based on a variety of different calibrations and suggest an age in the interval 4--8~Gyr.
This is, however, at the upper limit of where these relations have been calibrated.
Still, the fact that they indicate a relatively high age most likely excludes a low age of $\lesssim 3$~Gyr.
\paragraph*{This work}
The age estimates from this work differ significantly based on whether we fit to the magnitude or $\log g$.
When using the magnitude, the observed temperature is slightly cooler than the oldest isochrones, and we hit the upper edges of the grids.
However, when using $\log g$, we find a lower and well-defined age.
We see in the HR~diagrams that this is partly because the main sequences of the different isochrones are better separated in $\log g$ than in magnitude.
Since we use the value of $\log g$ determined from asteroseismology, we consider these age estimates to be the best we can get from isochrone fitting for this star.
Depending on the adopted isochrones, the $\log g$-based ages vary between 6--10~Gyr.
This star is also $\alpha$-enhanced, and when this is taken into account in the $Y^2$ isochrones, the age is lowered by 4~Gyr.
So, even though the precise value of $\log g$ is able to give a well-defined age estimate, it is very sensitive to the parameters of both the observables and the models.
\paragraph*{Conclusion}
Since the literature age based on asteroseismology does not pin down the age definitively, we include our own $\log g$-based ages and the ones based on rotation/activity in our definition of the age of this star: we give it as 4--10~Gyr, but the fits to $\alpha$-enhanced models indicate that the lower end of the interval may be more accurate.

\subsection*{$\alpha$ Cen B}
\paragraph*{Literature}
For this star, most of the literature ages are based on gyrochronology and a total of five different calibrations of the rotation-age relation.
Most of the estimates are consistent with each other in the range 4--7~Gyr; see the discussion of $\alpha$~Cen~A for a comment on the \citet{2011MNRAS.413.2218D} ages.
The two estimates based on two different calibrations of chromochronology agree with the gyrochronology ages.
The estimates based on model fitting also agree well with each other and the rest of the literature, except for the one by \citet{2007ApJS..168..297T} where only a lower limit is given.
Since they used the stellar parameters of \citet{2005ApJS..159..141V}, the age difference is most likely due to the use of different isochrones with a slightly shifted main sequence, but we cannot say with certainty.
The age given by \citet{2012MNRAS.425.3104F} of $5\pm1$~Gyr  may be the most reliable of the model fitting estimates since they also determined the age of $\alpha$~Cen~A, and discarded models for both stars which were too far removed from each other.
In that sense, their age estimate for $\alpha$~Cen~B is partially constrained to match the value of $\alpha$~Cen~A.
The one age based on asteroseismology is too low compared to the rest of the literature, the reason for this underestimation is discussed in \citet{2014A&A...566A..82L}.
More results with asteroseismology exist in the literature based on simultaneous modelling of $\alpha$ Cen A and B to give an age of the system.
Such estimates generally put the age of the system in the range 5--7~Gyr \citep{2002A&A...392L...9T, 2003A&A...402..293T, 2004A&A...417..235E, 2004ApJ...600..419G, 2005A&A...441..615M}.
\paragraph*{This work}
Our own age estimates differ slightly depending on whether we fit to the magnitude or $\log g$.
We consider the ages based on $\log g$ to be the most reliable since the isochrones, at the parameters of this star, have converged almost completely in luminosity.
This causes the magnitude-based ages to change with different isochrones due to slight changes in the main sequence.
The $\log g$-based ages show only a slight dependence on the adopted isochrones, and they are consistent with most of the literature values; however, the uncertainties extend all the way to the low edge of the grid in the case of $Y^2$ isochrones.
\paragraph*{Conclusion}
Given the large uncertainties on our own age estimates, and their sensitivity to the adopted isochrones and observables, we choose to base our conclusion for this star mainly on the gyrochronology and asteroseismic ages.
These age estimates show good consistency with the ones determined for $\alpha$~Cen~A and we choose to give the same range of 4--7~Gyr also for $\alpha$~Cen~B.

\subsection*{18 Sco}
\paragraph*{Literature}
This star has been the subject of many studies giving ages based on model fitting, and the different estimates lie in the range 2--6~Gyr.
The one outlier above 14~Gyr \citep{2012ApJ...746..101B} is based on a temperature (obtained from the interferometric radius) which is about 400~K lower than the typical literature values.
This moves the star towards the older isochrones and likely explains the high age they find.
This star is a solar twin, and the five most recent model fitting ages are based on very precise stellar parameters from differential spectroscopic analyses.
Considering just the estimates from these studies, the age is in the range 3--5~Gyr.
Since these studies find near identical stellar parameters, the remaining scatter is likely a reflection of a lower limit to the accuracy which is set by the differences between isochrones and model fitting algorithms.
The one age based on asteroseismology is slightly below 4~Gyr, consistent with the rest of the model fits.
Also the ages based on both gyrochronology and chromochronology agree well with the rest of the literature.
\paragraph*{This work}
Our own age estimates differ slightly depending on whether we fit to the magnitude or $\log g$.
We consider the ages based on $\log g$ to be the most reliable since the isochrones, at the parameters of this star, are less separated in luminosity.
This causes the magnitude-based ages to change with different isochrones due to slight changes in the location of the main sequence.
The $\log g$-based ages show only a slight dependence on the adopted isochrones, and they are completely consistent with the literature values.
\paragraph*{Conclusion}
Based on the most recent model fitting ages of this star, which are based on solar twin analyses, as well as the age estimates from gyrochronology and asteroseismology, we give the age as 3--5~Gyr.

\subsection*{$\mu$ Ara}
\paragraph*{Notes on input parameters}
The effective temperature determined by \citet{2015A&A...582A..49H} was not recommended for use as a reference value.
Instead, we adopted the mean spectroscopic literature value \citep[Table 11]{2015A&A...582A..49H}.
\paragraph*{Literature}
The literature ages based on model fitting are generally very precise and fall within the interval 4--8~Gyr, with the two estimates based on asteroseismology narrowing the interval down to 5--7~Gyr.
This star is located at the turn-off where the isochrones are starting to separate; this gives precise age determinations, but differences in isochrones and input parameters can still shift the age significantly.
There are also eight ages based on rotation/activity of which two are around 2~Gyr and the rest are at 6--8~Gyr.
The two low ages are both based on the calibration by \citet{1998MNRAS.298..332R}, who introduced a metallicity dependent correction to an earlier calibration, and are therefore strongly correlated.
The deviation of these age estimates from all the rest indicates a problem with their correction method.
The six higher values are based on three independent calibrations, but at these ages the rotation/activity-indicators are not necessarily reliable since none of the relations used here have been calibrated with stars older than the Sun.
Still, the fact that they indicate a relatively high age most likely excludes a low age of $\lesssim 3$~Gyr.
\paragraph*{This work}
Our age estimates fall in the lower end of the literature interval, and we see no significant difference between the use of different isochrones or input parameters.
The result using PARSEC isochrones and the magnitude has an upper uncertainty extending up to 8~Gyr; we find this solution to have a multimodal G-function.
This indicates that there is some ambiguity in the age determination of this star, and a slight change in input parameters may shift the solution towards a higher value.
\paragraph*{Conclusion}
Based mainly on the literature model fits and our own estimates, we give the age as 4--8~Gyr.
This may be slightly conservative given the two asteroseismic age estimates; however when their uncertainties are included they are close to being consistent with the entire interval of 4--8~Gyr.

\subsection*{$\beta$ Vir}
\paragraph*{Literature}
All of the literature values based on model fitting agree well on an age in the range 2--4~Gyr, with most of them falling near the middle of the interval.
Out of the seven estimates based on rotation/activity-age relations, one of the X-ray ages and the single rotation-based age fall within 2--4~Gyr.
The other X-ray age is underestimated, and the chromochronology ages are all higher than the ones based on model fitting.
They are also scattered which seems to be due to the use of different calibrations and activity measurements.
For example, \citet{2012AJ....143..135V} uses the calibration of \citet{2008ApJ...687.1264M} with the activity measurement by \citet{2004ApJS..152..261W} and find an age in between the ones given by \citeauthor{2008ApJ...687.1264M} and \citeauthor{2004ApJS..152..261W}.
\paragraph*{This work}
Our age estimates all agree very well across different isochrones and input parameters.
They are all very close to 3~Gyr in agreement with the isochrone ages in the literature.
\paragraph*{Conclusion}
The age of this star is well determined by isochrone fitting, and based on both our own values and the literature, we give the age as 2--4~Gyr.

\subsection*{Arcturus}
\paragraph*{Notes on input parameters}
We took $\log g$ from the recommendation given in the discussion in \citet[section~6.1]{2015A&A...582A..49H}.
\paragraph*{Literature}
The literature values are all based on model fitting, and most of them show good agreement around 6--10~Gyr.
However, \citet{2015ApJ...812...96G} and \citet{2011ApJ...743..135R} both used the same input data and the PARAM code to obtain their ages.
\citet{2008A&A...480...91S} also used the PARAM code (with the same isochrones) which means that most of the literature ages are somewhat correlated.
The lower of the two values determined by \citet{2011ApJ...743..135R} was based on a different fitting algorithm, and different isochrones, using the same data.
The lower age determined by \citet{2014MNRAS.443..698S} did not incorporate spectroscopic information; it is instead based on photometry and the distance.
\paragraph*{This work}
We see a clear difference between fitting to the magnitude or $\log g$.
The $\log g$ fits give lower ages, but the upper uncertainties stretch all the way to the upper edges of the grids since the $\mathcal{G}$~functions are essentially flat.
This is because the isochrones converge in surface gravity on the giant branch.
The ages based on the magnitude are more well-defined since the isochrones are better separated here.
We get ages which are consistent with the four highest literature values, but we find larger uncertainties which we can trace back to the uncertainties on our input parameters being larger.
When including $\alpha$-enhancement in the $Y^2$ isochrones, we find a significant shift towards a lower age.
This indicates that the MIST and PARSEC ages are overestimated.
\paragraph*{Conclusion}
With our $\alpha$-enhanced $Y^2$ isochrones we find a lower age limit of about 4~Gyr, and the literature suggests an upper limit of about 10~Gyr.
Thus, we give the age as 4--10~Gyr.

\subsection*{HD 122563}
\paragraph*{Literature}
For this star we have only found a single literature value of $9.7\pm2.9$~Gyr based on the Bayesian fit to photometry and spectra by \citet{2014MNRAS.443..698S}; however, they remark that this star has a bad photometric $T_{\mathrm{eff}}$.
\paragraph*{This work}
Our own ages all hit the upper edge of the grids since the star is cooler than the oldest isochrones at the observed metallicity (by about 400~K).
We do not know what causes this large discrepancy between the models and the observations, and it is too large to be solved by the inclusion of its observed $\alpha$-enhancement in the models.
\paragraph*{Conclusion}
Our own age estimates give no information, due to a discrepancy between the models and observations, and there is only one literature value.
Thus, we cannot reach a final conclusion on the age of this star.

\subsection*{$\mu$ Leo}
\paragraph*{Literature}
The three literature values are all based on model fitting and they are consistent within the uncertainties.
The two lower values \citep{2014A&A...566A..67L, 2018AJ....155...30B} are both based on the PARAM code using their own derived stellar parameters.
They adopt temperatures differing by 100~K which may explain the age difference.
\paragraph*{This work}
Our age estimates do not depend strongly on whether we fit to the magnitude or $\log g$; however, the latter gives less precise results due to the small separation of the isochrones as a function of $\log g$.
For the fits using the magnitude, the $Y^2$ isochrones give a lower and more precise value because these isochrones lack evolutionary stages beyond the RGB.
The ambiguity between this star being in the RC or RGB phase of evolution increases the age estimate and uncertainties when we use the MIST and PARSEC isochrones in which the clump phase is included (see the $\mathcal{G}$~functions in \autoref{fig:gfuncs_subset}c).
Our values are less precise than the two lowest literature values which seems to be due to a larger uncertainty on our metallicity (0.15~dex compared to 0.05~dex for both of the literature values) combined with slightly different values which moves the star to be slightly hotter than the RC.
\paragraph*{Conclusion}
If this star is on the RGB, its age is most likely 2--4~Gyr as found with the $Y^2$ isochrones and in the two lowest literature values.
However, the benchmark temperature and metallicity place it right on the RC in the HR~diagram which means that it may be older.
Based on this ambiguity, we give the age as 2--7~Gyr.

\subsection*{$\beta$ Gem}
\paragraph*{Literature}
All the literature ages are based on model fitting, and other than \citet{2018AJ....155...30B} agree on a low age of around 1~Gyr.
\citet{2018AJ....155...30B} derived their effective temperature based on their interferometric measurement of the stellar radius; its value is about 300--400~K lower than the spectroscopic estimates used to derive the other literature ages.
This difference is large enough to be the main reason for the age difference.
All but \citet{2015A&A...580A..24D} used the PARAM code to derive their ages, and the input parameters adopted by \citet{2016A&A...588A..98M}, \citet{2015ApJ...812...96G}, and \citet{2013A&A...554A..84M} are identical.
This means that most of the estimates are correlated.
\paragraph*{This work}
Our age estimates generally agree with the literature estimates on a value around 1~Gyr.
The one value which differs is based on fitting to the $Y^2$ isochrones using $\log g$.
The PDF of this fit has a peak at an age of 3~Gyr and a metallicity of [Fe/H] $\approx-0.1$~dex, which is lower than the benchmark metallicity of [Fe/H] $\approx0.15\pm0.16$~dex.
It is not clear to us why this solution is favoured over the younger and higher metallicity solution.
\paragraph*{Conclusion}
This is a young giant for which both the literature and our own values favour an age of around 1~Gyr.
This value is, however, quite sensitive to the input parameters; if the star is cooler or more metal-poor than what we assume, it is also older.
Based on the scatter in the age estimates, we give the age as 0.8--1.5~Gyr.

\subsection*{$\epsilon$ Vir}
\paragraph*{Literature}
All the literature ages are based on model fitting using different input data and algorithms, and they all agree on a low age of around 0.5--1~Gyr.
\paragraph*{This work}
Our age estimates fall within the range 0.5--1~Gyr in agreement with the literature values.
The age estimates are very precise since the isochrones are well separated for young giants, and the systematic uncertainty related to the use of different isochrones and input parameters is larger than the statistical uncertainty.
Two of the $\log g$-based ages are above 1~Gyr due to bimodal $\mathcal{G}$~functions caused by the overlap of a young RGB and an older AGB.
\paragraph*{Conclusion}
We choose to weigh the magnitude-based ages highest for this star and disregard the higher $\log g$-based ages.
Therefore, we give the age as 0.4--1.2~Gyr based on the scatter in our own magnitude-based results combined with the literature.

\subsection*{$\xi$ Hya}
\paragraph*{Literature}
The three literature ages are based on model fitting, and they all use different input data and fitting algorithms.
Two of them are in the range 0.5--1~Gyr, and the last one, by \citet{2014MNRAS.443..698S}, is at 4~Gyr.
\citeauthor{2014MNRAS.443..698S} state that their fit gives a questionable metallicity; they find an expectation value of the metallicity of [Fe/H] $=-0.46$~dex which is low compared to the benchmark value of [Fe/H] $=-0.16\pm0.20$~dex.
This difference in metallicity is likely enough to explain the age they find.
\paragraph*{This work}
Our age estimates fall within the range 0.5--1~Gyr in agreement with the literature values.
The age estimates are very precise since the isochrones are well separated for young giants, and the systematic uncertainty related to the use of different isochrones and input parameters is larger than the statistical uncertainty.
\paragraph*{Conclusion}
For a young giant like this, the age is well-defined based on isochrone fitting, and based on both our own values and those in the literature (except for the one by  \citet{2014MNRAS.443..698S} due to their questionable fit) we give the age as 0.5--1~Gyr.

\subsection*{HD 107328}
\paragraph*{Literature}
The two literature values are based on model fitting, and they both find values close to 7~Gyr.
However, considering the uncertainties, they are consistent with the range 4--10~Gyr.
\paragraph*{This work}
Both of the ages based on $Y^2$ isochrones are unreliable since those isochrones do not include models beyond the RGB, and this star is located where the RGB overlaps with more advanced evolutionary stages.
For the other two sets of isochrones, only the magnitude-based ages give any meaningful constraints since the isochrones converge in $\log g$ on the RGB.
Both magnitude-based ages are in the range 1--6~Gyr when including the uncertainties, but the $\mathcal{G}$~functions have extended tails which reach all the way up to the upper edge of the grid.
This makes the mean of the distribution quite different from the mode; in this case the mean of the $\mathcal{G}$~function is 6~Gyr which is closer to the literature values which both used the mean of the distribution instead of the mode.
The mode is at 3~Gyr for the PARSEC results since the star falls on the RGB of the 3~Gyr isochrones, and the RGB models are given a higher weight than the AGB models because of the longer evolutionary time.
The magnitude-based MIST $\mathcal{G}$~function shows a very narrow spike at an age just below 2~Gyr which is reduced in probability when fitting to the current surface metallicity instead of the initial.
So the difference in age between the current/initial metallicity fits are not present if we adopt the mean of $\mathcal{G}$~function as our age estimate instead of the mode.
\paragraph*{Conclusion}
The two literature values indicate an age in the interval 4--10~Gyr, and our best estimates are lower at 1--6~Gyr.
The difference is essentially due to our estimates being based on the mode of the G-function instead of the mean.
This choice has a large impact in this case because the G-function has an extended tail towards high ages due to the ambiguity between it being in the RGB phase of evolution or beyond.
Without more information about the evolutionary state of this star, we give the age as 1--10~Gyr.

\subsection*{HD 220009}
\paragraph*{Notes on input parameters}
The effective temperature and $\log g$ determined by \citet{2015A&A...582A..49H} were not recommended for use as reference values.
Instead, we adopted the mean spectroscopic literature values \citep[Table 11]{2015A&A...582A..49H}.
\paragraph*{Literature}
For this star we have only found a single literature value of $6.2\pm3.9$~Gyr based on the Bayesian fit to photometry (without spectroscopic data) by  \citet{2014MNRAS.443..698S}.
\paragraph*{This work}
This star has the largest uncertainty on $\log g$ in this sample (0.34~dex), and it is located on the cool side of the oldest isochrone at its observed metallicity.
This, combined with the small separation of the isochrones, gives us no age information based on $\log g$.
When using the magnitude instead, the star is located within the isochrone grid and we find best estimates of the age ranging from about 4--8~Gyr depending on the isochrones.
However, the estimates are very uncertain (the $\mathcal{G}$~functions are almost completely flat) and the 1$\sigma$ intervals are consistent with an age in the range 2--13~Gyr.
The large uncertainties are a combination of the close spacing of the isochrones in this region of the HR~diagram, and the uncertainty on the effective temperature (111~K) which is the highest in this sample.
\paragraph*{Conclusion}
This star is located on the giant branch in the HR~diagram where the isochrones are poorly separated.
Combined with relatively high uncertainties in the stellar parameters, this means that none of our estimates give good constraints on the age.
Based on the $\mathcal{G}$~functions of our magnitude-based fits, we give the age as $>2$~Gyr.

\subsection*{$\alpha$ Tau}
\paragraph*{Literature}
For this star we have only found a single literature value of $5.9\pm3.8$~Gyr based on the Bayesian fit to photometry and spectra by \citet{2014MNRAS.443..698S}.
\paragraph*{This work}
This far up the giant branch there is almost no age information in $\log g$ since the isochrones converge.
Using the magnitude instead, the isochrones are better separated, and this star falls close to the old edge of the grid.
Still, the $\mathcal{G}$~functions are flat above 8~Gyr and extend down to about 4~Gyr at the lower end.
\paragraph*{Conclusion}
Our own magnitude-based ages imply that this star is older than 4~Gyr; however, the literature value extends down to 2~Gyr.
It is very difficult to get a precise age for a star like this, even with precise input parameters.
In the end, we simply give the age as  $>2$~Gyr.

\subsection*{$\alpha$ Cet}
\paragraph*{Literature}
For this star we have only found a single literature value of $5.5\pm3.7$~Gyr based on the Bayesian fit to photometry and spectra by \citet{2014MNRAS.443..698S}.
\paragraph*{This work}
This far up the giant branch there is almost no age information in $\log g$ since the isochrones converge.
Using the magnitude instead, the isochrones are better separated, and this star falls among the younger isochrones.
The magnitude-based ages are in the range 1--5~Gyr when including the uncertainties, but the $\mathcal{G}$~functions have extended tails which reach all the way up to the upper edge of the grid.
This is mainly due to the large uncertainty on the metallicity for this star (0.47~dex).
The extended $\mathcal{G}$~function makes the mean of the distribution quite different from the mode; in this case the mean of the distribution is 6.5~Gyr which is closer to the literature value which used the mean instead of the mode.
\paragraph*{Conclusion}
The literature value indicate an age in the interval 2--10~Gyr, and our best estimates prefer the low end of the interval, namely 1--5~Gyr.
The difference can be entirely explained by our estimates being based on the mode of the $\mathcal{G}$~function instead of the mean.
This choice has a large impact in this case because the $\mathcal{G}$~function has an extended tail towards high ages due to the large uncertainty on the metallicity.
Based mainly on our own results, we give the age as 1--10~Gyr.

\subsection*{$\beta$ Ara}
\paragraph*{Literature}
The two literature values are both based on model fitting, but they give very different ages.
\citet{2011MNRAS.410..190T} fit to the temperature and luminosity, assuming solar metallicity, and find
an age of 50~Myr.
\citet{2014MNRAS.443..698S} find an age of around 3~Gyr, but they did not use any spectroscopic data for this star.
\paragraph*{This work}
This is the most massive star in the sample (the benchmark value is 8.21~$M_\odot$ \citep{2015A&A...582A..49H}), and as a result it falls outside of our isochrone grids.
Therefore, we created a new grid of MIST isochrones with ages in the range 10--500~Myr in steps of 10~Myr as described in Section \ref{sec:special_beta_ara}.
Fitting to this grid with the $V$~magnitude, we find an age of $50\pm10$~Myr in agreement with the low literature value.
\paragraph*{Conclusion}
This is a relatively massive, young giant, and our estimate based on a grid of low ages is 50~Myr which is in agreement with the lowest of the two literature values.
Thus, we give the age as 40--60~Myr (0.04--0.06~Gyr).

\subsection*{$\gamma$ Sge}
\paragraph*{Literature}
For this star we have found two literature values based on model fitting which agree on an age of around 5~Gyr, but with uncertainties which in the worst case span the range 2--10~Gyr.
\paragraph*{This work}
This far up the RGB there is almost no age information in $\log g$ since the isochrones converge.
Using the magnitude instead, the isochrones are better separated, and this star falls among the younger isochrones.
The magnitude-based ages are in the range 1--4~Gyr, but with upper ends of the confidence intervals reaching 8~Gyr.
Additionally, the $\mathcal{G}$~functions have extended tails which reach all the way up to the upper edge of the grid due to the large uncertainty on the metallicity for this star (0.39~dex).
The extended $\mathcal{G}$~function makes the mean of the distribution quite different from the mode; in this case the mean of the distribution is 6.5~Gyr which is closer to the literature values which both used the mean instead of the mode.
\paragraph*{Conclusion}
The literature value indicates an age in the interval 2--10~Gyr, and our best estimates prefer the low end of the interval, namely 1--4~Gyr, but with large uncertainties.
The difference can be entirely explained by our estimates being based on the mode of the $\mathcal{G}$~function instead of the mean.
This choice has a large impact in this case because the $\mathcal{G}$~function has an extended tail towards high ages due to the large uncertainty on the metallicity.
Based mainly on our own results, we give the age as 1--10~Gyr.

\subsection*{$\psi$ Phe}
\paragraph*{Literature}
For this star we have only found a single literature value of $4.9\pm4.5$~Gyr based on the Bayesian fit to photometry by  \citet{2014MNRAS.443..698S}.
However, they note that their solution is outside their model grid, making the result unreliable.
\paragraph*{This work}
This star falls far off the cool edge of our isochrone grids for the adopted metallicity, so none of our age estimates can be considered to be reliable.
It would take a change in metallicity of about 1~dex to bring the models and observations to agree, and we do not know the source of this discrepancy.
\paragraph*{Conclusion}
We have obtained no reliable age estimates for this star and cannot reach a final conclusion on its age.

\subsection*{$\epsilon$ Eri}
\paragraph*{Literature}
We have found an almost equal number of ages based on model fitting and rotation/activity-age relations.
All ages based on rotation or activity fall within the range 0.4--0.9~Gyr, and they are based on four different calibrations of chromochronology, and three different calibrations of gyrochronology.
Adding the fact that this is an age at which the rotation/activity diagnostics are reliable, this is a strong indication that this star is younger than 1~Gyr.
The estimates based on model fitting, on the other hand, give a range of values up to about 10~Gyr.
This simply reflects the fact that isochrone dating is unreliable for low-mass dwarfs (which is also indicated by the large uncertainties).
\paragraph*{This work}
Since this star is on the main sequence, our ages are not very reliable.
Using the magnitude, we get values ranging from the lower edge of the grid up to about 10~Gyr.
The isochrones are slightly better separated using $\log g$, but even with an uncertainty as small as 0.03~dex, we find values of up to 6~Gyr which is significantly higher than the age of  less than 1~Gyr implied in the literature.
At the very least, the fit to $\log g$ is able to exclude an age much above 6~Gyr.
\paragraph*{Conclusion}
The isochrone-based ages using $\log g$ allow us to say that this star is likely young, but they have large uncertainties since this is a low-mass dwarf.
We consider the literature values based on rotation/activity diagnostics to be the most reliable for this star and give the age as 0.4--0.9~Gyr.

\subsection*{Gmb 1830}
\paragraph*{Notes on input parameters}
The effective temperature determined by \citet{2015A&A...582A..49H} was not recommended for use as a reference value.
Instead, we adopted the mean spectroscopic literature value \citep[Table 11]{2015A&A...582A..49H}.
\paragraph*{Literature}
We have found an almost equal number of ages based on model fitting and rotation/activity-age relations.
Most of the ages based on rotation or activity fall within the range 3--6~Gyr, and they are based on three different calibrations of chromochronology and two different calibrations of gyrochronology.
These methods are thought to be reliable in this age range, but they have not been calibrated for stars as metal-poor as this one.
The estimate given by \citet{1998MNRAS.298..332R} is based on their own metallicity-dependent correction to a previous calibration and gives a much higher age.
The estimates based on model fitting give a range of values spanning almost the entire possible range with a slight preference for values above 5~Gyr.
This simply reflects the fact that isochrone dating is unreliable for low-mass dwarfs.
\paragraph*{This work}
This star is slightly cooler than the main sequences of our isochrone grids which means that we find a most likely age which is at the upper edge of the grid in most cases.
However, the isochrones are so closely spaced that this provides no reliable information about the age.
\paragraph*{Conclusion}
All isochrone-based ages are unreliable for this star since it is a low-mass dwarf, so we are left with the ages based on rotation/activity.
However, at the metallicity of this star the reliability of these estimates is unknown and we cannot reach
a final conclusion on the age.

\subsection*{61 Cyg A}
\paragraph*{Literature}
The literature age values based on model fitting show some degree of agreement around an age of 6~Gyr; however, \citet{2007ApJS..168..297T} only give an upper limit, and \citet{2012ApJ...756...46R} find a value of 2~Gyr with high precision.
This high precision estimate is quite surprising given that this star is far enough down the main sequence that the isochrones have converged completely.
The age given by \citet{2008A&A...488..667K} is based on a fit including constraints on the radius (from their own interferometric observations) and the mass (based on radial velocity monitoring of the system; \citealt{1995Icar..116..359W}).
\citet{2014ApJ...780..159E} took the mass from the best fit by \citet{2008A&A...488..667K} and used it to rederive the age with a different set of stellar models.
So these two values are not completely independent, but they may represent the best available age estimates based on stellar models.
The four rotation/activity-age estimates are based on independent calibrations and give values in the range 1--4~Gyr.
At the lower end of this range, such estimates are thought to be reliable; however, the scatter is quite large for a reliable indicator (see e.g. $\epsilon$ Eri for an example with little scatter).
So there is a slight tension between the rotation/activity-based ages and the best model fitting ages, and it is not clear to us which are better.
\paragraph*{This work}
At this position on the main sequence, the $\mathcal{G}$~functions are almost completely flat, and the age estimates are therefore not reliable.
We find that the $Y^2$ isochrones give larger lower age limits simply because the main sequence is shifted slightly compared to the other isochrones.
\paragraph*{Conclusion}
The best estimate based on model fitting is an age of around 6~Gyr as found in the literature when radius and mass estimates are included in the fit.
However, the literature values based on rotation/activity diagnostics put this star at an age in the interval 1--4~Gyr.
It is not clear to us which of these values is the better estimate, so we give the age as 1--7~Gyr.
This is in good agreement with what we find for its binary companion 61 Cyg B.

\subsection*{61 Cyg B}
\paragraph*{Literature}
The two literature values based on model fitting which give an actual age estimate agree on a value of around 6--8~Gyr.
The age given by \citet{2008A&A...488..667K} is based on a fit including constraints on the radius (from their own interferometric observations) and the mass (based on radial velocity monitoring of the system; \citealt{1995Icar..116..359W}).
\citet{2014ApJ...780..159E} took the mass from the best fit by  \citet{2008A&A...488..667K} and used it to rederive the age with a different set of stellar models.
So these two values are not completely independent, but they may represent the best available age estimates based on stellar models.
The four rotation/activity-age estimates are based on independent calibrations and give values in the range 2--5~Gyr.
At the lower end of this range, such estimates are thought to be reliable; however, the scatter is quite large for a reliable indicator (see e.g. $\epsilon$ Eri for an example with little scatter).
So there is a slight tension between the rotation/activity-based ages and the best model fitting ages, and it is not clear to us which are better.
\paragraph*{This work}
At this position on the main sequence, the $\mathcal{G}$~functions are almost completely flat, and the age estimates are therefore not reliable.
The magnitude-based age we find using the PARSEC isochrones may seem well determined, but this is only due to the $\mathcal{G}$~function showing oscillating behaviour.
We believe this is due to the limited mass resolution of the isochrone grids on the main sequence combined with the low uncertainties on the stellar parameters.
\paragraph*{Conclusion}
The best estimate based on model fitting is an age of around 6--8~Gyr as found in the literature when radius and mass estimates are included in the fit.
However, the literature values based on rotation/activity diagnostics put this star at an age in the interval 2--5~Gyr.
It is not clear to us which of these values is the better estimate.
This is similar to what we find for its binary companion 61~Cyg~A, and we choose to give the same age range of 1--7~Gyr.


\bsp	
\label{lastpage}
\onecolumn
\includepdf[pages=-]{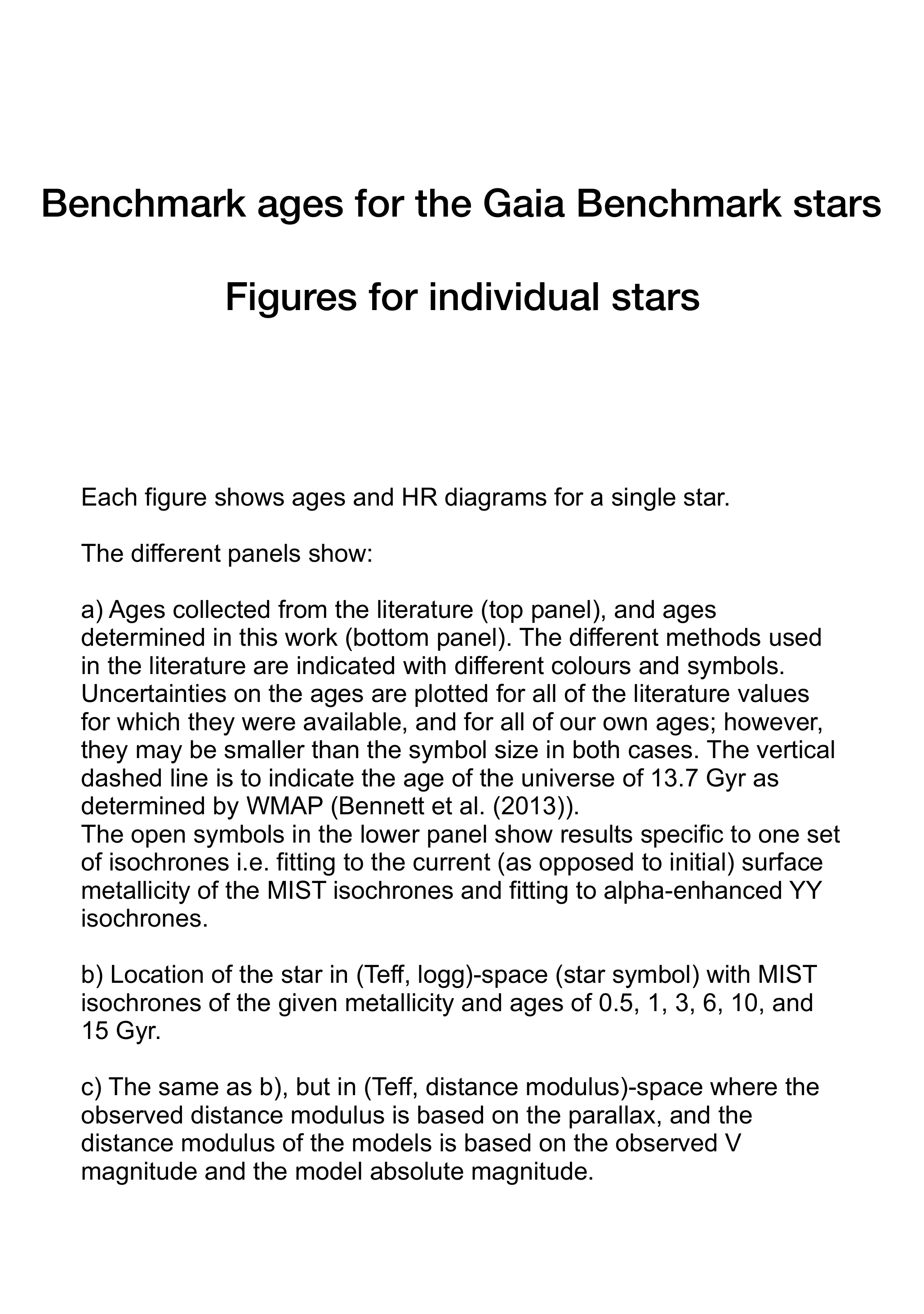}
\end{document}